\documentclass[aip,cha,amsmath,amssymb,reprint,numerical]{revtex4-1}

\usepackage{graphicx}
\usepackage{dcolumn}
\usepackage{bm}
\usepackage{subfigure}
\usepackage{color}
\usepackage[normalem]{ulem}
\usepackage{verbatim}
\usepackage{epsfig}

\begin{document}

\title{Temporal organization of magnetospheric fluctuations unveiled by recurrence patterns in the Dst index}

\author{Reik V. Donner}
\affiliation{Research Domain IV -- Transdisciplinary Concepts and Methods, Potsdam Institute for Climate Impact Research, Potsdam, Germany}

\author{Veronika Stolbova}
\affiliation{Research Domain IV -- Transdisciplinary Concepts and Methods, Potsdam Institute for Climate Impact Research, Potsdam, Germany}
\affiliation{Department of Physics, Humboldt University, Berlin, Germany}
\affiliation{Department of Banking and Finance, University of Zurich, Switzerland}
\affiliation{Department of Control Theory, Nizhny Novgorod State University, Nizhny Novgorod, Russia}

\author{Georgios Balasis}
\affiliation{Institute for Astronomy, Astrophysics, Space Applications and Remote Sensing, National Observatory of Athens, Penteli, Athens, Greece}

\author{Jonathan F. Donges}
\affiliation{Research Domain I -- Earth System Analysis, Potsdam Institute for Climate Impact Research, Potsdam, Germany}
\affiliation{Stockholm Resilience Centre, Stockholm University, Stockholm, Sweden}

\author{Marina Georgiou}
\affiliation{Institute for Astronomy, Astrophysics, Space Applications and Remote Sensing, National Observatory of Athens, Penteli, Athens, Greece}
\affiliation{Department of Physics, Section of Astrophysics, Astronomy and Mechanics, National and Kapodistrian University of Athens, Zografos, Athens, Greece}
\affiliation{Mullard Space Science Laboratory, Department of Space \& Climate Physics, University College London, Dorking, Surrey, United Kingdom}

\author{Stelios M. Potirakis}
\affiliation{Department of Electronics Engineering, Piraeus University of Applied Sciences (TEI Piraeus), Aigaleo, Athens, Greece}

\author{J\"urgen Kurths}
\affiliation{Research Domain IV -- Transdisciplinary Concepts and Methods, Potsdam Institute for Climate Impact Research, Potsdam, Germany}
\affiliation{Department of Physics, Humboldt University, Berlin, Germany}
\affiliation{Department of Control Theory, Nizhny Novgorod State University, Nizhny Novgorod, Russia}
\affiliation{Institute for Complex Systems and Mathematical Biology, University of Aberdeen, Aberdeen, United Kingdom}

\date{\today}

\begin{abstract}
Magnetic storms constitute the most remarkable large-scale phenomena of nonlinear magnetospheric dynamics. Studying the dynamical organization of macroscopic variability in terms of geomagnetic activity index data by means of complexity measures provides a promising approach for identifying the underlying processes and associated time-scales. Here, we apply a suite of characteristics from recurrence quantification analysis (RQA) and recurrence network analysis (RNA) in order to unveil some key nonlinear features of the hourly Disturbance storm-time (Dst) index during periods with magnetic storms and such of normal variability. Our results demonstrate that recurrence-based measures can serve as excellent tracers for changes in the dynamical complexity along non-stationary records of geomagnetic activity. In particular, trapping time (characterizing the typical length of ``laminar phases'' in the observed dynamics) and recurrence network transitivity (associated with the number of the system's effective dynamical degrees of freedom) allow for a very good discrimination between magnetic storm and quiescence phases. In general, some RQA and RNA characteristics distinguish between storm and non-storm times equally well or even better than other previously considered nonlinear characteristics like Hurst exponent or symbolic dynamics based entropy concepts. Our results point to future potentials of recurrence characteristics for unveiling temporal changes in the dynamical complexity of the magnetosphere.
\end{abstract}


\maketitle

\begin{quotation}

Geomagnetic activity indices trace the temporal variability of the Earth's magnetic field across different spatial domains of the near-Earth environment. Among others, the Dst index has been used in many previous studies as a diagnostic of the overall state of the magnetosphere. Here, we employ different techniques based upon the concept of recurrence plots to improve our understanding of the complex variability patterns exhibited by this index when the geomagnetic field undergoes a sequence of magnetic storm and quiescience periods triggered by non-stationary solar wind forcing. We demonstrate that recurrence characteristics provide unique tools for discriminating between the dynamical complexity properties of Dst during times with strong geomagnetoc activity and quiescence phases. Thus, future applications of these measures to other geomagnetic activity indices with higher temporal resolution may potentially allow us to identify characteristic signatures of complexity variations preceeding intense magnetic storms, which could open new perspectives for space weather short-term forecasting.

\end{quotation}

\section{Introduction}

Various effects of space weather present a natural hazard to which the modern human civilization has become increasingly vulnerable, particularly through the use of ever more sophisticated technologies (see \cite{Song2011} and references therein). The magnetosphere of the Earth is generally a benign host for satellite communication and global positioning systems, but can change into a quite inhospitable environment. The miniaturization of electronic components with which spacecrafts are equipped renders them susceptible to damage by charged particles accelerated to high energies by impulsive geomagnetic field disturbances \cite{Singer2010}. On the surface of the Earth, electrical currents induced during geospace storms, when absorbed, can damage long-line power networks connecting large geographic regions \cite{Boteler1998,Bolduc2002,Wik2009}.

Geospace magnetic storms occur as a disturbance in the Earth's magnetic field, driven by large-scale eruptions of plasma and magnetic fields from the solar corona launched onto a trajectory that impacts the Earth's magnetosphere \cite{Bothmer2007,Richardson2012}. Similar to severe weather phenomena, as well as other types of natural disasters associated with the Earth's internal dynamics like earthquakes and volcanic eruptions, they vary remarkably in the severity of disturbance. However, unlike the latter events, geospace storms have a global reach, and their effects can be seen simultaneously around the Earth. The solar wind provides a continuous input in the form of mass, momentum and energy. If not dissipated, it is stored in the magnetotail until, through a sequence of energy-loading and stress-developing processes, the magnetospheric system is reconfigured \cite{Baker2007}. During magnetic storms, charged particles in the radiation belts are accelerated to high energies resulting in intensified electric current systems causing characteristic signatures in the Earth's magnetic field \cite{Baker2005,Daglis2008}. 

As a consequence of the solar wind forcing, the magnetosphere is continuously far from equilibrium and undergoes complex variations at a broad range of temporal and spatial scales \cite{Consolini2008}. Its response is typically not proportional to the forcing and commonly changes abruptly rather than gradually. Taken together, there is strong evidence to consider the magnetosphere as a complex system with distinct, nonlinearly coupled regions, where multiple interlinked phenomena occur on a vast range of length and time scales \cite{Watkins2001}. Since the first pieces of evidence of large-scale coherence provided by observations of low-dimensional behavior in time series of auroral electrojet indices \cite{Vassiliadis1990}, studying magnetospheric activity dynamics from the viewpoint of nonlinear dynamics and complexity has gained substantial new insights into the response of the magnetosphere to solar wind energy input. 

Taking our understanding of the temporal variations of the magnetic field at the surface of the Earth and in the surrounding space a step further, additional evidence of a hierarchical multi-scale organization of magnetospheric activity has been found in the form of characteristic scaling laws in some dynamical properties of regional and global geomagnetic activity indices \cite{Takalo1994,Wei2004}. By making use of a  phase transition approach \cite{Shao2003,Sharma2006}, global coherency was reconciled with scale-invariance. Specifically, it was demonstrated that some global features exhibit properties typical for phase transitions of first order, while the multi-scale properties appear compatible with second-order transitions \cite{Sitnov2001,Balasis2006}. 

In this work, we utilize the powerful framework of recurrence analysis for studying some additional nonlinear properties of the Earth's magnetosphere. The fundamental idea of recurrence analysis is based on the long-known fact that many natural processes obey a distinct recurrent behavior in time, ranging from very regular diurnal or annual variability in meteorological variables over almost periodic oscillatory patterns (such as Milankovich cycles in the Earth's climate history or the return intervals of extrema of cosmic-ray intensity measured at the surface of the Earth \cite{Monk1939}) to more irregular climatic modes such as the El Ni\~{n}o Southern Oscillation. From a dynamical systems perspective, the recurrence of states (i.e., finding new states arbitrarily close to previously assumed states if waiting long enough) is a fundamental property of both deterministic and stochastic dynamics \cite{Marwan2007,Marwan2014}. Recently, it has been proven mathematically that the temporal pattern of such recurrences allows for reconstructing the dynamics of the underlying variable up to a monotonous transformation, which implies that recurrences contain fundamental information about the dynamical organization of the system under study \cite{Robinson2009}.

Here, we utilize a suite of measures from recurrence quantification analysis (RQA) \cite{Marwan2007} and recurrence network analysis (RNA) \cite{Donner2011} to study the nonlinear dynamics exhibited by hourly values of the disturbance storm-time (Dst) index (i.e., the average change of the horizontal component of the Earth's magnetic field recorded at four mid-latitude magnetic observatories) during one year of observations. Specifically, we consider the year 2001, since it showed several distinct phases of geomagnetic activity \cite{Balasis2008,Balasis2009}. Both RQA and RNA have recently proven useful for quantifying dynamical complexity in non-stationary models as well as real-world time series \cite{Bastos2011,Zbilut2002, Schinkel2009}, including applications to geoscientific problems \cite{Donges2011,Donges2011PNAS}. Here, both frameworks are combined to study the complex signatures of magnetospheric fluctuations during non-storm and storm conditions as reflected in the Dst index. Our aim is to unveil non-trivial features of magnetospheric dynamics originating from the time-dependent coupling between different nonlinear subsystems, particularly such properties that are not captured by other linear and nonlinear methods of time series analysis. However, by focusing on a single geomagnetic activity index only, the primary goal of this work is to establish the applicability of quantitative recurrence-based characteristics in the context of geomagnetism and space weather, while detailed studies on a larger set of relevant variables allowing to obtain more detailed results and interpretations on specific geophysical processes and mechanisms are outlined as a subject of future work.

This paper is organized as follows: Section~\ref{sec:method} describes the data and methods used in this study in some detail. The temporal variability of the dynamical complexity of the Dst index as uncovered by various recurrence-based characteristics is reported in Section~\ref{sec:performance}. Subsequently, we discuss the capability of the different measures to discriminate between storm and quiescence periods with respect to a heuristic global-scale and a data-adaptive local-scale classification based on the Dst index in Sections~\ref{sec:discrimination} and \ref{sec:adaptive}, respectively. Implications of our results are addressed in Section~\ref{sec:discussion}.

\section{Data and methods} \label{sec:method}

\subsection{Description of the data} \label{sec:data}

The previous solar cycle 23 (May 1996 to January 2008), peaked in 2000 -- 2003 with many strong solar events, which in turn caused extended periods of strong magnetospheric activity. In particular, the year 2001 saw the occurrence of two intense magnetic storms on 31 March 2001 and 6 November 2001 (when Dst reached minimum values of $-387$~nT and $-292$~nT, respectively), which were associated with two large coronal mass ejections (CMEs) on 29 March 2001 and 4 November 2001, respectively.

\begin{figure}
\centering
\resizebox{0.45\textwidth}{!}{\includegraphics*{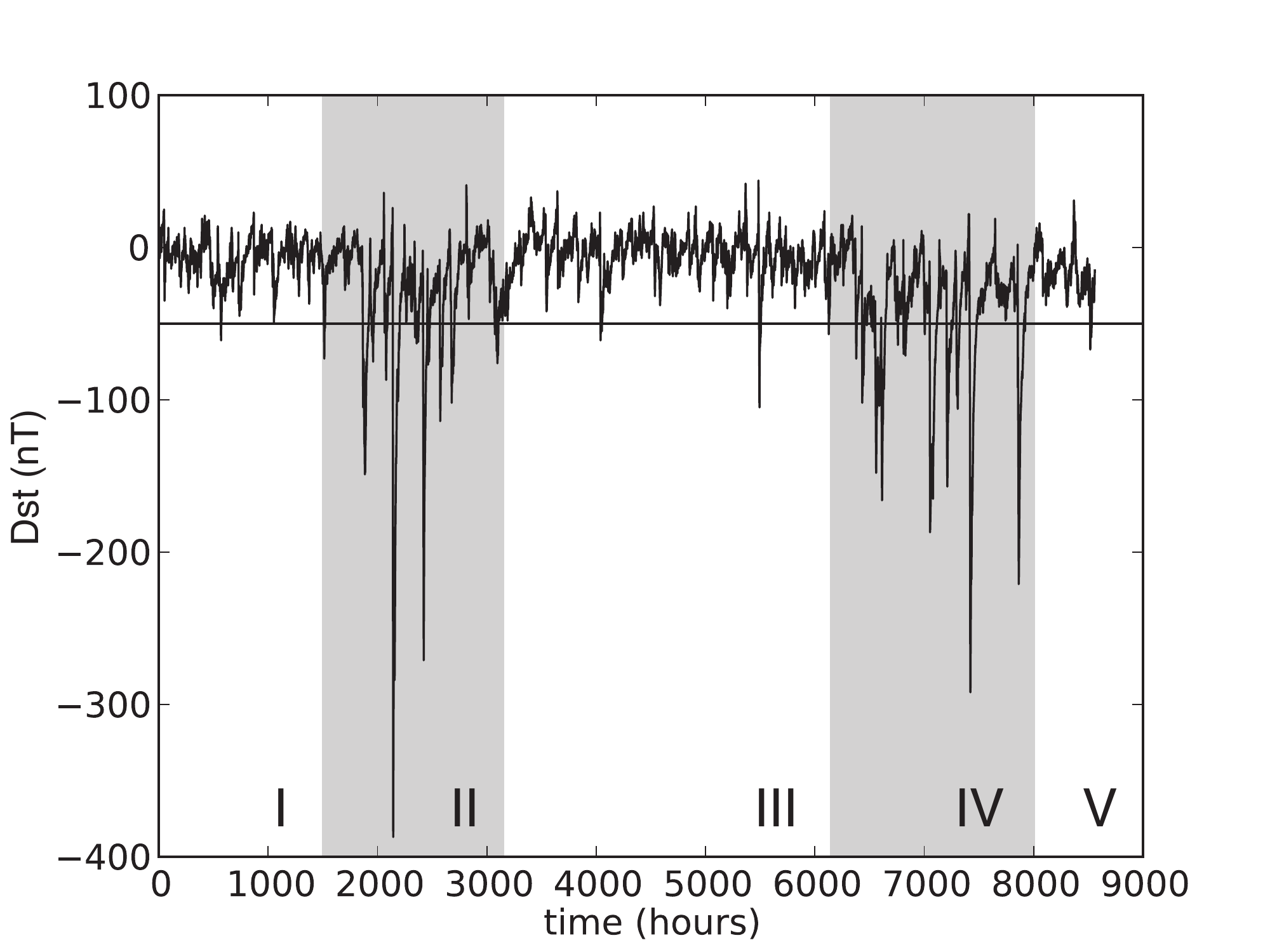}}
\caption{Dst index time series for the period from 1 January 2001 to 31 January 2002. Gray shaded areas highlight the two time intervals (II, IV) heuristically classified as storm periods, whereas the remaining periods (I, III, V) are considered as episodes of quiescence. The horizontal line depicts a value of Dst$=-50$~nT, which is commonly considered as a criterion for defining a magnetic storm.}
\label{fig:data}
\end{figure}

Figure~\ref{fig:data} shows the corresponding Dst time series. It can be seen that the data can be rather naturally divided into 5 shorter time series (I-V in Fig.~\ref{fig:data}). The second and fourth time windows (II and IV in Fig.~\ref{fig:data}, highlighted by gray shading) include the Dst variations related to the periods around the aforementioned intense magnetic storms of March and November 2001, respectively. The same time intervals were previously shown to be compatible with the emergence of two distinct patterns in the Earth's magnetosphere at time scales between hours and a few weeks (i.e., which are relatively long in comparison to the short-term bursty dynamics of the magnetosphere): (i) a pattern associated with intense magnetic storms (``pathological'' states of the magnetosphere) in windows II and IV, which have been characterized by a higher degree of organization, and (ii) a pattern associated with ``normal'' (non-storm) periods (``physiological'' states of the magnetosphere) in windows I, III and V, which are characterized by a lower degree of organization, i.e., more random fluctuations \cite{Balasis2006,Balasis2008,Balasis2009,Balasis2011a,Donner2013}. 

\begin{figure}
\centering
\resizebox{0.485\textwidth}{!}{\includegraphics*{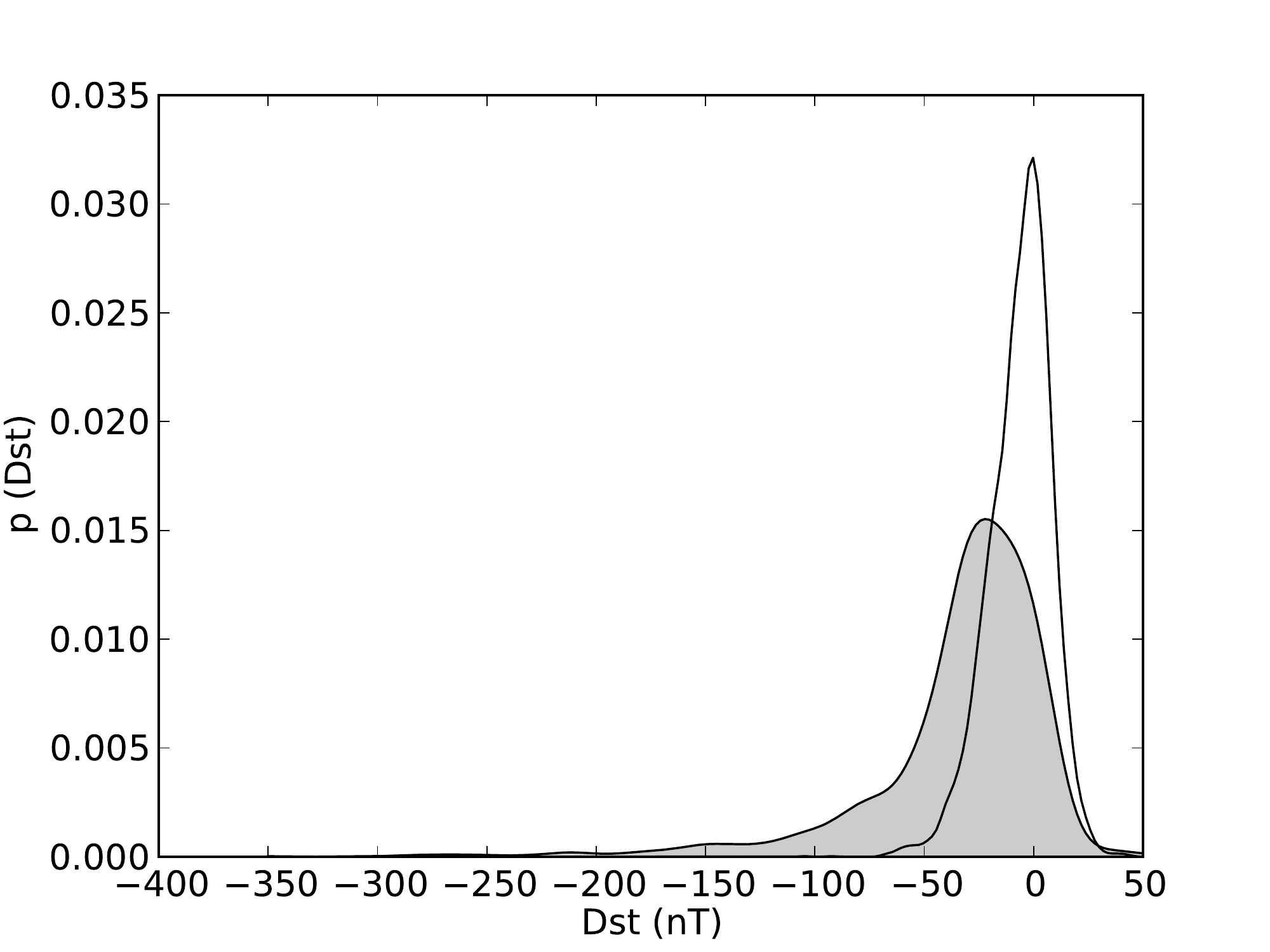}}
\caption{Probability density functions (Gaussian kernel estimates) of hourly Dst values during storm (gray area) and quiescence (white area) periods as considered in this paper (see text).}
\label{fig:pdf_dst}
\end{figure}

We note that this heuristic global distinction into storm and non-storm periods does not lead to a clear separation between all the hourly Dst values recorded within the different periods. As Fig.~\ref{fig:pdf_dst} shows, there is a considerable overlap between the probability density functions (PDFs) of hourly Dst values observed during both types of periods, which is mainly due to the fact that the actual individual storms (strongly negative Dst values) have a much shorter duration than the considered overall storm periods (note that geomagnetic storms typically appear clustered in time), so that there are a lot of close to zero Dst values even during storm periods. In turn, there have also been some weaker storms during the supposed non-storm periods. This observation does not contradict our paradigm of separating the time series under study into distinct periods, since the dynamical characteristics to be studied will be computed for running windows instead of individual points in time (see below). In this respect, we still expect that these characteristics (relating to time-scales from days to weeks) exhibit marked differences between time periods prone to severe magnetic storms and those with relative quiescence of the magnetosphere.

\subsection{Time-delay embedding}

The methods to be used in the following are based on the concept of phase space in deterministic dynamical systems, which is commonly spanned by a multitude of complementary variables that -- taken together -- uniquely determine the system's state. In our case, however, we have access to only a single variable (Dst), the dynamics of which contains contributions due to both, the ``internal'' dynamics of the geomagnetic field and the external forcing by the solar wind and related extra-terrestrial processes, which are mutually entangled in some complex, nonlinear way. In such a case, where one only observes a one-dimensional projection of the system's ``true'' physical phase space, it can be hard (or even practically impossible) to infer meaningful information from the data comprising only a single variable (i.e., applying classical univariate time series analysis), since the actually relevant dynamics might take place orthogonal to this ``coordinate''.

\begin{figure*}[t!]
\centering
\resizebox{0.95\textwidth}{!}{\includegraphics*{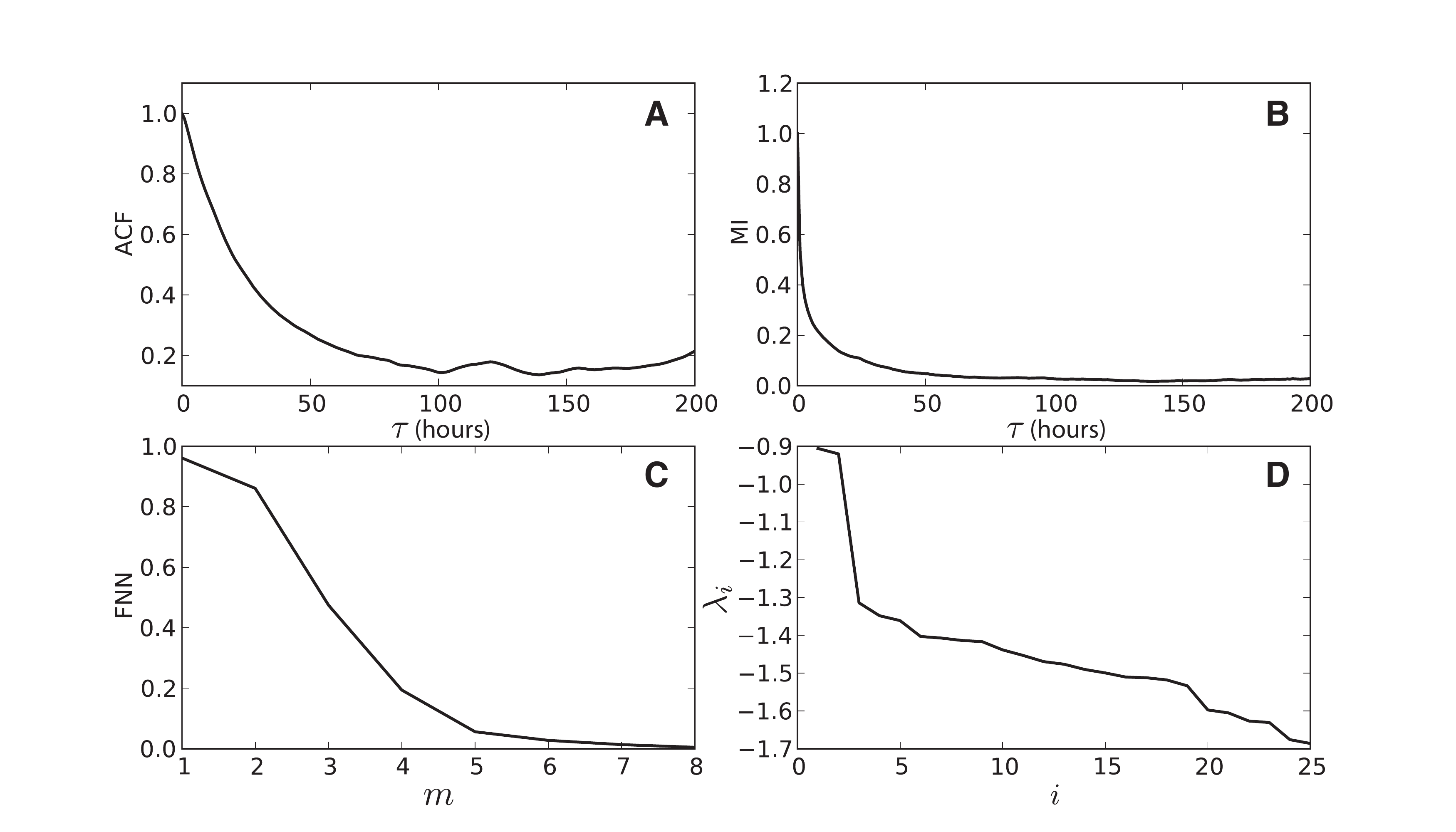}}
\caption{Determination of embedding parameters for the Dst index time series: (A) auto-correlation function, (B) mutual information, (C) number of false nearest-neighbors (FNN) in dependence on the embedding dimension $m$, (D) logarithm of the leading normalized eigenvalues ($\lambda_i=\log_{10}(\sigma_i^2 / \sum_i \sigma_i^2)$) of the lagged-trajectory covariance matrix using a delay of $\tau=100$  as suggested by panels (A) and (B) and $m=25$ time-shifted replications of the data.}
\label{fig:embedding}
\end{figure*}

One way to circumventing the aforementioned problem is reconstructing the phase space from its single observed coordinate by making use of possible signatures originating from the influences of other variables. If we presume the observed dynamics to be deterministic and dissipative, we can apply a theoretical framework developed by \cite{Mane1981} and \cite{Takens1981} to qualitatively reconstruct unobserved variables from our time series by means of time-delay embedding \cite{Packard1980,Takens1981}. For this purpose, we construct a multivariate time series $X(t)$ from the original data $x(t)$ by considering appropriate time-shifts of the data:
\begin{equation}
\begin{split}
X(t)&=(x(t),x(t-\tau),x(t-2\tau),\dots,x(t-(m-1)\tau)) \\
&=(X^{(1)}(t),\dots,X^{(m)}(t)).
\label{eq:embedding}.
\end{split}
\end{equation}
\noindent
For selecting the number $m$ and delay $\tau$ of such time-shifted copies of our data, several approaches have been proposed and utilized in a wide range of applications. Specifically, a reasonable choice of the embedding delay $\tau$ can commonly be obtained by considering estimates of the de-correlation time, e.g., the delay associated with the first zero-crossing or the first decay below 1/e of the auto-correlation function (ACF), or the first minimum of the time-averaged mutual information (MI) \cite{Fraser1986}. An appropriate choice of the embedding dimension $m$ can be inferred by means of the false nearest-neighbor method \cite{Kennel1992}, where changes in the proximity relations between state vectors in the reconstructed phase space are monitored as the embedding dimension is increased. Such changes typically arise due to projection effects if the number of components considered is not yet sufficient. An alternative approach is studying the scaling of eigenvalues of the covariance matrix of $X(t)$ for very high-dimensional embeddings, which should reveal the number of dynamically relevant variables -- an approach closely related to singular spectrum analysis \cite{Broomhead1986}.

Figure~\ref{fig:embedding} shows the results of the aforementioned methods for the Dst index time series of the year 2001. Here, we intentionally disregard the known non-stationarity of Dst due to the alternation between storm and non-storm periods, as also seen in classical time series properties like power spectra studied in previous works \cite{Balasis2006} and manifested here in the very slow decay of the ACF in Fig.~\ref{fig:embedding}A. Instead, we apply all aforementioned methods to the complete time series, thereby effectively integrating information from different time scales from relatively short-term (hourly to sub-daily) fluctuations to the slow succession of activity and quiescence periods. If doing so, both auto-correlation and mutual information function decay within at most a couple of days to values clearly below 1/e, cf. Fig.~\ref{fig:embedding}A,B. Since there is some ambiguity among the different rules of thumb for choosing a proper embedding delay, we take $\tau=100$ hours in the following, which accounts for the fact that during storm periods, we may expect the emergence of long-range correlations and, thus, a slower decay of serial dependencies \cite{Balasis2006,Donner2013}. As we will demonstrate later in Section~\ref{sec:settings}, varying the embedding delay over some reasonable range of values (for which the individual components $X^{(k)}$ ($k=1,\dots,m$) of the embedding vector are sufficiently uncorrelated) does not alter the results of our recurrence analysis, which is consistent with findings previously made in other geoscientific applications \cite{Donges2011}. 

Regarding the embedding dimension $m$, the two methods mentioned above suggest somewhat different choices (Fig.~\ref{fig:embedding}C,D): while the number of false nearest neighbors decays to values close to zero only at about $m\approx 5$ to $7$, the spectrum of the lagged-trajectory covariance matrix shows a marked drop already after the second-largest eigenvalue. As a compromise between both results, and taking the requirements of a windowed analysis with sufficiently high temporal resolution (i.e. small window size, see below for details) into account, we will use $m=3$ in all further computations. This choice represents a trade-off between a probably higher dimensionality of the dynamical system represented by the Dst index and the temporal resolution of the data in conjunction with the time scales of interest (up to a few days) to be resolved by our approach.

\subsection{Recurrence plots}

Recurrence plots have been originally introduced as a simple means to visualizing the temporal succession of close states in phase space \cite{Marwan2007,Eckmann1987}. Given a time series containing the relevant system variables (either from multivariate observations or time-delay embedding of univariate signals), we define the recurrence matrix $\mathbf{R}=(R_{ij})$ as a binary matrix encoding whether or not each pair of state vectors on the sampled trajectory is mutually close. One possible formulation, which we will adopt in this work, is making this decision by comparing all pairwise distances with some threshold distance $\varepsilon$,
\begin{equation}
R_{ij}(\varepsilon)=\Theta(\varepsilon-\|X(t_i)-X(t_j)\|_\infty),
\end{equation}
\noindent
where $\Theta(\cdot)$ denotes the Heaviside function and $\|\cdot\|_\infty$ the maximum norm defined as
\begin{equation}
\|X(t_i)-X(t_j)\|_\infty=\max_k \left|X^{(k)}(t_i)-X^{(k)}(t_j)\right|.
\end{equation}

\begin{figure}
\centering
\resizebox{0.45\textwidth}{!}{\includegraphics*{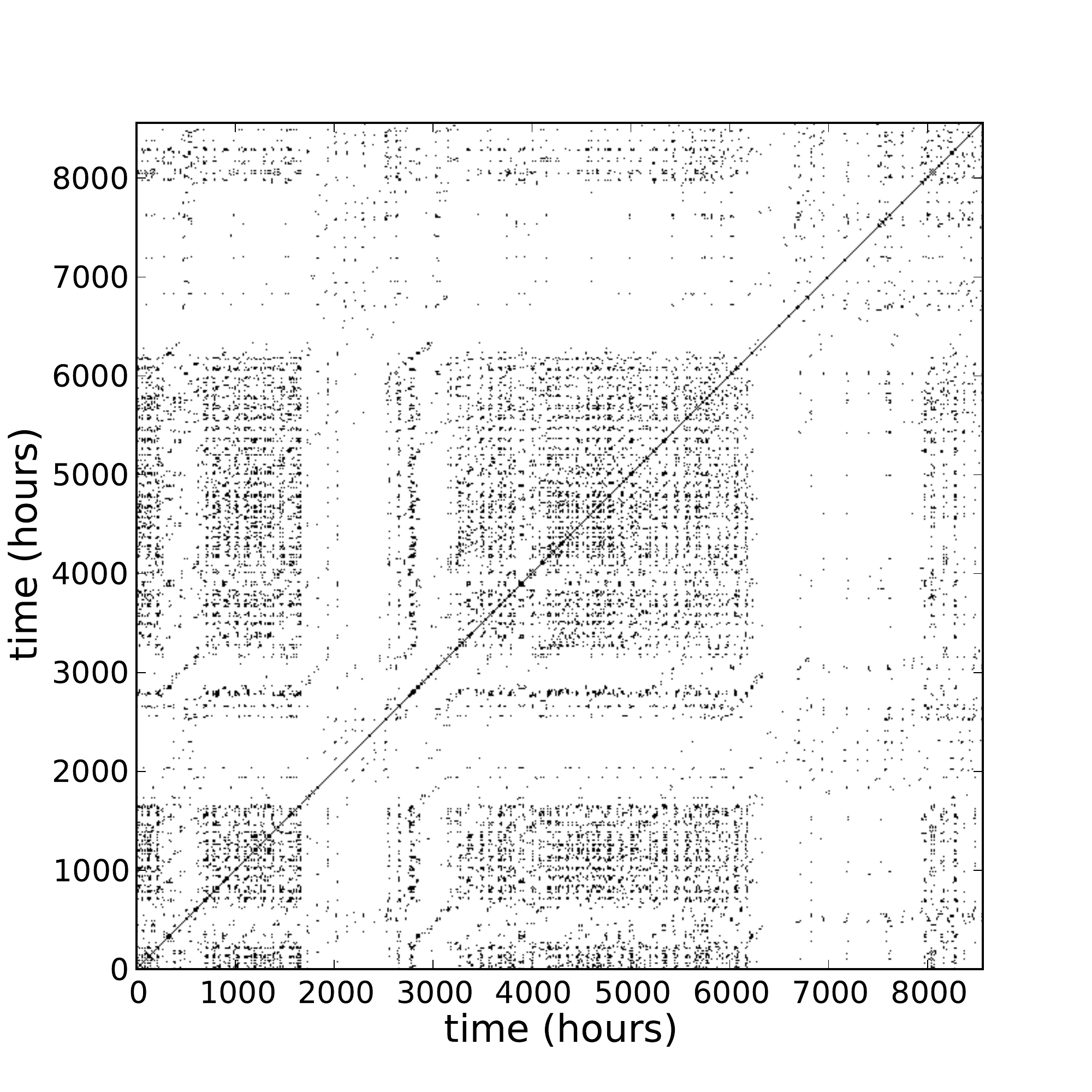}}
\caption{Recurrence plot for the Dst index time series computed with the embedding parameters $m=3$ and $\tau=100$, using a global recurrence rate of $RR=0.05$.}
\label{fig:rp}
\end{figure}

In Fig.~\ref{fig:rp}, we display the recurrence plot of the Dst data (i.e., the visual representation of the binary recurrence matrix $\mathbf{R}$). For illustrative purposes, we use here a recurrence threshold $\varepsilon$ chosen such as to yield a recurrence rate of $RR=0.05$, i.e., the 5\% closest pairs of state vectors are considered as recurrences. (Since the Dst data are given as integer values, we add some small Gaussian white noise with standard deviation of 0.01 before computing the distances to avoid problems with properly fixing the recurrence rate. It has been checked that different realizations or somewhat different standard deviations of the noise do not affect the results presented in this paper.) One immediately recognizes that the density of recurrences is much higher during non-storm periods than during storm times, resulting in some distinct block structure of the recurrence plot. This is mainly due to the large negative Dst amplitudes associated with magnetic storms. Specifically, when considering the same absolute differences in Dst values for defining the closeness of two values during storm and quiescence periods, it is easy to see that storm conditions (characterized by a larger variance of Dst values than periods without magnetic storms) tend to exhibit a lower number of recurrences. In a similar way, non-stationary systems exhibiting a trend would show a subsequent loss of recurrences at larger temporal distances between observations (i.e., fewer recurrences in the upper left and lower right part of the recurrence plot). However, as revealed by Fig.~\ref{fig:rp}, this kind of behavior is less relevant for the observed dynamics of the Dst index, since we do not observe any systematic trend in the Dst data but rather a succession of time intervals with low and high amplitudes manifested in the obtained block structure.

In order to unveil changes in the dynamical characteristics of the Dst index during storm and non-storm periods, in the following we will consider sliding windows in time (with width $w$ and mutual offset $\Delta w$, see below). For obtaining a corresponding quantitative characterization of the time-dependent recurrence properties, different methodological settings would be possible. On the one hand, we might keep the global recurrence threshold $\varepsilon$ fixed and take slighting windows along the main diagonal of the recurrence plot in Fig.~\ref{fig:rp} to quantify changes in the contained recurrence structures. In this case, the recurrence rate would become a function of time, taking high values in the periods of magnetospheric quiescence and lower ones in the time intervals characterized by magnetic storms. Since some of the characteristic measures discussed below exhibit explicit or implicit dependencies on the fraction of recurrences, we will not further consider this approach. Instead, we will fix the recurrence rate $RR=0.05$ and compute recurrence plots individually for each time window with this value of $RR$, thereby allowing for some quantitative comparison between properties calculated for different time slices. This strategy causes $\varepsilon$ to become a function of time itself. However, the variability of this parameter is exclusively determined by the (multivariate) distribution of (time-lagged) Dst values in the reconstructed phase space and, thus, has only limited additional value. As an alternative, a variety of more complex statistical characteristics can be estimated from the recurrence structures within the individual time windows. Here, we restrict our attention to two particularly useful frameworks, recurrence quantification analysis (RQA) and recurrence network analysis (RNA), both of which provide a variety of measures capturing different  dynamical and geometric properties of the underlying data.

\subsection{Recurrence quantification analysis (RQA)}

RQA is a widely used method of nonlinear time series analysis \cite{Marwan2007,Marwan2014}, which statistically characterizes the frequency and duration of episodes of recurrences of a dynamical system represented by a single trajectory in phase space. More specifically, RQA is based on the idea that deterministic dynamics gives rise to the emergence of line structures in recurrence plots. Notably, diagonal lines in the recurrence plot  formed by recurrent states only (i.e., off-diagonal structures in the recurrence matrix containing exclusively values of 1) correspond to a similar evolution of the system within two time slices \cite{Zbilut1992,Webber1994}. The longer such diagonal lines, the more predictable is the observed dynamics. For an ideal periodic dynamics, the recurrence plot would exclusively consist of such non-interrupted diagonal lines. In contrast, diagonal structures of a certain minimum length can hardly be found in stochastic systems, since mutually close sequences of states arise here only due to chance, becoming increasingly unlikely as the considered sequence length increases. For a complex yet deterministic system, we can expect that recurrences always tend to form diagonal line structures instead of arising individually at random. This idea provides the basis for the so-called ``degree of \textit{determinism}'' \cite{Marwan2007},
\begin{equation}
DET = \frac{\sum_{d=d_{min}}^{d_{max}} d\ p(d)}{\sum_{d=1}^{d_{max}} d\ p(d)},
\end{equation}
\noindent
where $d$ denotes the length of a diagonal line in the recurrence plot, $p(d)$ gives the associated probability density function, $d_{max}$ is the maximal diagonal line length (excluding the main diagonal in the plot, referred to as the line of identity), and $d_{min}\geq 2$ (we use $d_{min}=2$ in what follows). Specifically, if all recurrences belong to diagonal line structures, this implies $DET=1$, indicating a maximal degree of determinism of the system's dynamics.

In addition to diagonal line structures, vertical lines (and, due to the symmetry of the recurrence plot, also horizontal lines) in the recurrence plot emerge whenever the system's state does not change much within a certain time window \cite{Marwan2002}. Following a similar rationale as for the diagonal lines discussed above, the fraction of recurrences contained in such vertical structures thus provides a measure for the degree of \textit{laminarity} of the dynamics,
\begin{equation}
LAM = \frac{\sum_{v=v_{min}}^{v_{max}} v\ p(v)}{\sum_{v=1}^{v_{max}} v\ p(v)},
\end{equation}
\noindent
where $v$ denotes the length of a vertical line and $p(v)$ the associated probability density. The mean length of such structures is referred to as the \textit{trapping time}
\begin{equation}
TT = \frac{\sum_{v=v_{min}}^{v_{max}} v\ p(v)}{\sum_{v=v_{min}}^{v_{max}} p(v)}.
\end{equation}
\noindent
In both cases, we will use $v_{min}=2$ in the following.

Although there is a large variety of other complementary RQA measures \cite{Marwan2007}, in this work we will restrict our attention to the aforementioned three characteristics. One important reason for this selection is that for providing a discrimination between the dynamics of the Dst index during storm and non-storm periods, we have observed that these measures commonly work better than other RQA characteristics (not shown). Moreover, they belong to those RQA measures that have been most widely applied to data from various scientific disciplines in the past and mostly have some rather intuitive interpretation \cite{Marwan2007}.

In order to test for the presence of episodes of extraordinary dynamical characteristics, we adopt a randomization procedure \cite{Schinkel2008}, which consists of determining the expected distribution of RQA measures from bootstrapping the respective line length distribution from the recurrence plot obtained from the entire time series. In contrast to the original approach, in which only lines with a distance of less than $w$ from the main diagonal ($i=j$) are chosen (corresponding to the running window size of our analysis), we bootstrap here from the line distributions taken from the entire recurrence plot. The rationale for this modification is that we consider the system as being in some quasi-stationary state during most of the times for which we observe a high number of recurrences (i.e., mainly non-storm periods). Thus, by taking the entire recurrence plot into account, we increase the sample size of line structures used for bootstrapping. In turn, due to the differences in the local recurrence rates during storm and non-storm periods when taking the same value of $\varepsilon$, the applied testing procedure does not fully account for the fundamental differences in the dynamical state of the magnetosphere during both types of periods, but exhibits an intrinsic bias towards non-storm characteristics. As a consequence, we cannot expect that only the RQA measures during storm periods deviate significantly from the distribution obtained by bootstrapping, since the latter are based on a mixture of two line length distributions associated with storm and non-storm periods, respectively. In this spirit, we do not consider our approach as a significance test as proposed by \cite{Schinkel2008}, but as providing a baseline for expected dynamical properties when taking information from randomly chosen time intervals into account.

\subsection{Recurrence network analysis (RNA)}

In the last years, there have been considerable efforts in exploring different transformations from time series into complex networks \cite{Donner2011,Zhang2006,Xu2008,Marwan2009,Donner2010,Donges2013EPL}, which allow using versatile tools from complex network theory for characterizing different aspects related to the complex organization of trajectories of dynamical systems. Among others, some first successful applications of such approaches to studying the dynamics of solar activity have been reported recently \cite{Yu2012,Zou2014NJP,Zou2014NPG}.

One of the so far most popular approaches of complex network-based time series analysis is RNA, which directly builds upon the recurrence plot concept. RNA makes use of the fact that the state vectors of the system are arranged in some metric space \cite{Donner2011,Marwan2009,Donner2010}. Disregarding the main diagonal, the proximity relations between these state vectors can thus be considered as representing the edges of some random geometric graph. For this purpose, we identify each state vector with a node and define the adjacency (connectivity) matrix of the associated graph as
\begin{equation}
A_{ij}(\varepsilon)=R_{ij}(\varepsilon)-\delta_{ij},
\end{equation}
\noindent
where $\delta_{ij}$ denotes the Kronecker delta. The asymptotic properties of this recurrence network are (for sufficiently large sample size) exclusively determined by the geometric structure of the underlying system in phase space \cite{Donner2011EPJB,Donges2012}. Based on this idea, one can use complex network characteristics for quantifying structural (as opposed to dynamical) properties of the system under study \cite{Donner2011,Donner2010}.

According to our previous experience in studying dynamical transitions from paradigmatic model systems as well as real-world geoscientific time series \cite{Donges2011,Donges2011PNAS,Marwan2009,Zou2010,Donges2014}, we select the following three quantifiers that characterize the resulting recurrence networks at a global scale:

\begin{itemize}

\item \textit{Network transitivity} measures the fraction of connected triples of nodes (having at least two edges) that are fully connected (i.e., exhibit all three possible edges and are thus referred to as ``triangles''):
\begin{equation}
\mathcal{T}= \frac{\sum_{i,j,k} A_{ij} A_{ik} A_{jk}}{\sum_{i,j,k} A_{ij} A_{ik}}.
\end{equation}
\noindent
(Note that we have omitted the $\varepsilon$-dependence of $A_{ij}$ for brevity.)
As shown by \cite{Donner2011EPJB}, $\mathcal{T}$ has some intuitive interpretation in terms of the effective dimension of the system, i.e., its number of degrees of freedom. High transitivity points to a very regular (low-dimensional) dynamics, whereas low values indicate more complex fluctuations.

\item The \textit{global clustering} coefficient provides a conceptually related measure, quantifying the mean transitivity of the respective subgraphs that contain a given node $i$ of the network together with its neighbors $\mathcal{N}(i)=\{j:A_{ij}=1\}$. Specifically,
\begin{equation}
\mathcal{C}=\frac{1}{N}\sum_i \frac{\sum_{j,k} A_{ij} A_{ik} A_{jk}}{\sum_{j,k} A_{ij} A_{ik}}.
\end{equation}
\noindent
Depending on the connectivity distribution among the network's nodes, $\mathcal{C}$ can exhibit an entirely different behavior than $\mathcal{T}$ and, thus, provide complementary information. Relating to the interpretation of $\mathcal{T}$, we may consider the global clustering coefficient as a proxy for the mean local dimensionality of the system \cite{Donner2011EPJB}.

\item Finally, we utilize the \textit{average path length} $\mathcal{L}$, i.e., the mean minimum number of edges separating all pairs of nodes in the recurrence network from each other. As shown recently \cite{Donges2011,Donges2011PNAS,Zou2010}, $\mathcal{L}$ is particularly sensitive to qualitative changes in the dynamics of the system under study, and may thus be considered as an indicator of abrupt dynamical changes.

\end{itemize}

Similar as for the RQA measures, we are interested in whether or not the aforementioned RNA characteristics deviate significantly from the properties of a recurrence network sampled from state vectors belonging to arbitrary points in time. This is achieved by randomly drawing individual samples from the full set of state vectors corresponding to the entire Dst time series and creating recurrence networks from these random samples with the same overall recurrence rate as in our windowed analysis \cite{Donges2011}. Note that this approach is distinctively different from that used for the RQA measures, since the linkage relations are not fixed before drawing the random sample (as in case of line bootstrapping with fixed $RR$), but are generated after bootstrapping the individual state vectors. In this spirit, the test provided by this procedure has an entirely different meaning than the line bootstrapping test in RQA \cite{Schinkel2008}. For the purpose of this study, we will consider it as a test against stationarity: assuming the system being on average in some stationary state, deviations of the RNA parameters from the distributions taken from the bootstrapped samples arise from some ``non-typical'' temporary geometric alignment of nodes in the reconstructed phase space, which could in fact appear during both, storm and non-storm periods.

\section{Results}\label{sec:performance}

\subsection{Temporal variations of recurrence characteristics}

Following the strategy described in Section~\ref{sec:method}, we perform a recurrence analysis with a fixed $RR=0.05$ in all windows, $\tau=100$ hours and $m=3$. For displaying the results, we use a slightly modified convention of the embedding vectors differing from Eq.~(\ref{eq:embedding}) by taking replications shifted towards the future instead of the past, i.e., 
\begin{equation}
X(t)=(x(t),x(t+\tau),x(t+2\tau)).
\end{equation}
\noindent
This modification only fixes the reference time scale to the beginning of each considered time window (instead of its end) and does not mean that future values are used for characterizing the present state of the magnetosphere.

\begin{figure}
\centering
\resizebox{0.485\textwidth}{!}{\includegraphics*{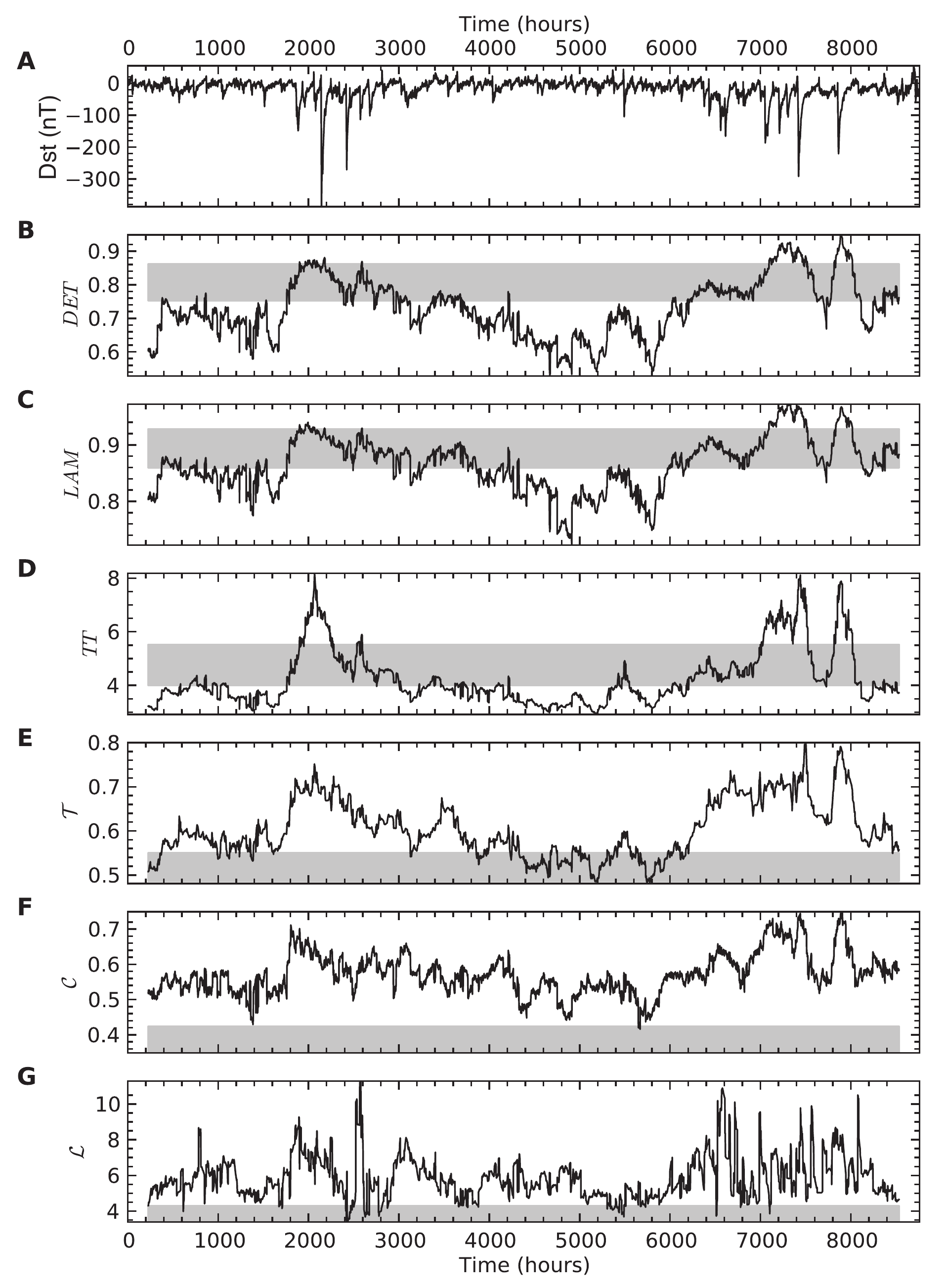}}
\caption{Dst time series (A) and RQA and RNA measures for the year 2001: (B) determinism $DET$, (C) laminarity $LAM$, (D) trapping time $TT$, (E) transitivity $\mathcal{T}$, (F) global clustering coefficient $\mathcal{C}$, (G) average path length $\mathcal{L}$ (computed for running windows of size $w=256$ hours and mutual offset $\Delta w=1$ hour). Gray shades indicate the expected 90\% range provided by the bootstrapping approaches introduced in Section~\ref{sec:method}. Note that the tests for RQA and RNA measures are based on different rationales, so the outcomes have essentially different interpretations (see Section~\ref{sec:bootstrapping} for some more detailed discussion). All recurrence measures are displayed according to the contemporary component $x(t)$ of the embedding vector Eq.~(\ref{eq:embedding}) at the corresponding window midpoint.}
\label{fig:results}
\end{figure}

The results for a representative window size of $w=256$ hours (i.e., about 10 days, capturing the essential time scales of magnetic storms, cf.~\cite{Donner2013,Johnson2005}) are shown in Fig.~\ref{fig:results}. According to the relatively large embedding delay, this implies that we effectively take information from Dst values into account that spread over a considerably larger time window of $w+(m-1)\tau=456$ hours. By doing so, we utilize variations on time scales between hours and a few weeks for our analysis, disregarding any higher or lower frequency variability. As can be seen, both RQA and RNA characteristics exhibit marked temporal variability corresponding to the alternation between storm and non-storm periods. Specifically, during storm periods, all six considered measures take distinct maxima. For the RQA parameters $DET$, $LAM$ and $TT$ as well as the RNA measures $\mathcal{T}$ and $\mathcal{C}$, this is consistent with a higher degree of organization in the system as proposed earlier \cite{Balasis2006}, leading to a more regular (and thus more predictable) dynamics. In the Dst index, this behavior is typically reflected by gradual trends corresponding to the emergence and subsequent disappearance of the magnetic field perturbation over a time scale comparable to the considered window size \cite{Donner2013}. In turn, the maxima of $\mathcal{L}$ during storm periods are rather indicative of persistent qualitative changes of dynamical complexity, which trace the succession of individual storms interrupted by short periods of quiescence in magnetospheric fluctuations during these activity phases.

With the exception of $\mathcal{L}$, the considered measures characterize related but conceptually different aspects of dynamical complexity of the geomagnetic field fluctuations captured by Dst. In particular, $DET$ and $LAM$ show an extremely strong similarity. This indicates that in the Dst index, periods of higher degree of dynamical organization (i.e., ``smoother'' variability with respect to the full temporary range of states, commonly emerging during storm periods) coincide with a tendency towards more laminar dynamics (i.e., values ``not changing much'' over certain periods of time). The latter are probably related to generally weaker fluctuations superimposed to the gradual Dst trends during magnetic storms, given that the adaptive recurrence threshold $\varepsilon$ follows the general variance of Dst within the considered time intervals. In other words, we observe a separation of ``normal'' background fluctuations from larger-scale temporary trends in the Dst index, and both $DET$ and $LAM$ appear to trace primarily the variation of the level of detail provided by our methods.

In a similar way, we observe that $TT$ and $\mathcal{T}$ present closely related temporal variability profiles. We interpret this observation as both characteristics being associated with the regularity of fluctuations, one ($TT$) capturing this property from the dynamical, the other ($\mathcal{T}$) from a geometric perspective. We particularly observe maxima of both measures during storm periods, which are more distinct than in case of $DET$ and $LAM$, suggesting that $TT$ and $\mathcal{T}$ are more sensitive tracers of dynamical changes in the magnetospheric variability during storm phases than $DET$ or $LAM$. 

Moreover, we find that also $\mathcal{T}$ and $\mathcal{C}$ present a similar evolution, with $\mathcal{C}$ exhibiting less distinct maxima. The latter behavior is not uncommon in applications of RNA, since both properties are based on similar considerations. However, by construction $\mathcal{C}$ exhibits some sensitivity to changes in the heterogeneity of connectivity distributions in the recurrence networks, whereas $\mathcal{T}$ does not. Recalling the close relationship of $\mathcal{T}$ with a generalized notion of fractal dimension \cite{Donner2011EPJB}, it is not surprising that the transitivity presents some more distinct variability profile sensitively tracing changes in the degree of regularity of the observed dynamics.

Notably, in comparison with the other five measures, $\mathcal{L}$ shows a less clear yet still visible signature of transitions between storm and non-storm periods. As mentioned above, we relate this observation to the fact that this measure primarily traces the succession of different dynamical ``states'' (individual magnetic storms interrupted by short periods of quiescence) during storm periods consisting of several events, while the remaining two RNA measures as well as the three RQA characteristics quantify the overall complexity of fluctuations.

\begin{table}
\centering
\begin{tabular}{|c|cccccc|}
\hline
& $DET$ & $LAM$ & $TT$ & $\mathcal{T}$ & $\mathcal{C}$ & $\mathcal{L}$ \\
\hline
$DET$ & 
1.00 & 0.97 & 0.90 & 0.89 & 0.85 & 0.50 \\
$LAM$ & 
0.97 & 1.00 & 0.85 & 0.86 & 0.87 & 0.47 \\
$TT$ & 
0.90 & 0.85 & 1.00 & 0.87 & 0.80 & 0.49 \\
$\mathcal{T}$ & 
0.89 & 0.86 & 0.87 & 1.00 & 0.82 & 0.58 \\
$\mathcal{C}$ & 
0.85 & 0.87 & 0.80 & 0.82 & 1.00 & 0.53 \\ 
$\mathcal{L}$ & 
0.50 & 0.47 & 0.49 & 0.58 & 0.53 & 1.00 \\
\hline
\end{tabular}
\vspace{0.2cm}
\caption{Pearson's correlation coefficient between the 
different recurrence parameters for a window size of $w=256$ hours, offset $\Delta w=1$ hour.}
\label{tab:correlations}
\end{table}

The aforementioned interdependences are confirmed and further quantified by Tab.~\ref{tab:correlations} displaying Pearson's correlation coefficients \cite{Barlow1989} between the individual measures 
taken over the respective running windows. For the sake of simplicity, we restrict this analysis to the consideration of linear correlations between the different measures, ignoring possibly more complex (nonlinear) interrelations as well as effects due to strong deviations of the associated probability distribution functions from Gaussianity.

\subsection{Amplitude-complexity relationship}

The qualitative assessment provided above points to a close relationship between the amplitude of the Dst index and the complexity of its temporal variations. Notably, this aspect goes beyond classical mean-variance interrelationships often observed in heteroscedastic signals (for which Dst might be a good example), but describes a specific feature of magnetospheric dynamics. In the following, we further quantify these interrelationships between recurrence characteristics and the window-wise mean Dst index.

\begin{table}
\centering
\begin{tabular}{|c|ccc|}
\hline
& PC($0$) & PC($\tau$) & PC($2\tau$) \\
\hline
$DET$ & -0.74 & -0.76 & -0.71 \\
$LAM$ & -0.69 & -0.72 & -0.69 \\
$TT$ & -0.77 & -0.80 & -0.74 \\
$\mathcal{T}$ & -0.78 & -0.81 & -0.75 \\
$\mathcal{C}$ & -0.63 & -0.70 & -0.65 \\ 
$\mathcal{L}$ & -0.47 & -0.53 & -0.48 \\
\hline
\end{tabular}
\vspace{0.2cm}
\caption{Pearson's correlation coefficient (PC) between the window-wise mean hourly Dst index and the different recurrence parameters for a window size of $w=256$ hours, offset $\Delta w=1$ hour. For defining the running windows for the recurrence-based measures, the first, second and third components of the embedding vectors ($x(t)$, $x(t+\tau)$ and $x(t+2\tau)$, respectively) have been considered.}
\label{tab:correlations2}
\end{table}

As already expected from Fig.~\ref{fig:results}, Tab.~\ref{tab:correlations2} reveals a general anti-correlation between all measures and the mean Dst index in each window, which underlines the fact that maxima of the considered recurrence characteristics commonly coincide with magnetic storms (strongly negative Dst values). Specifically, $\mathcal{T}$ and $TT$ generally exhibit the strongest linear (anti-) correlations with Dst, followed by $DET$, $LAM$ and $\mathcal{C}$, whereas $\mathcal{L}$ shows significantly weaker correlations with the Dst index values. This finding allows formulating a preliminary hypothesis regarding the appropriateness of different recurrence parameters for tracing signatures of magnetic storms. Specifically, we expect that $\mathcal{T}$ and $TT$ allow better discriminating the dynamical state of the magnetosphere during storm and non-storm periods than the other measures. We will further investigate this hypothesis in Sections~\ref{sec:discrimination} and \ref{sec:adaptive}. Note again, that this simple linear correlation analysis disregards any effects due to non-normality and possible nonlinearity of the relationships between the different recurrence properties and Dst.

\begin{figure}
\centering
\resizebox{0.485\textwidth}{!}{\includegraphics*{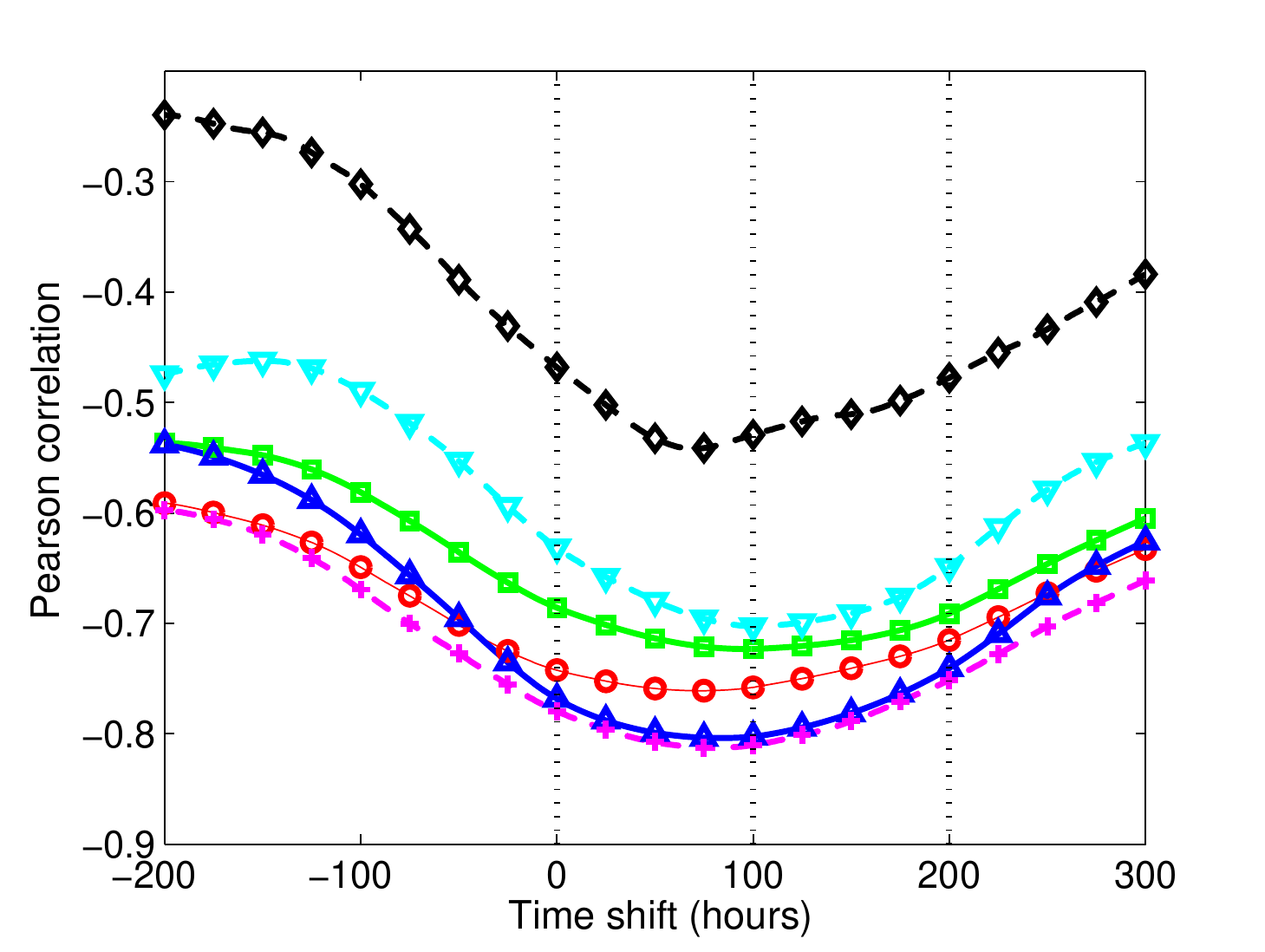}}
\caption{(Pearson) cross-correlation functions between the window-wise mean Dst time series on the one hand, and the RQA measures (solid) $DET$ (red, $\bigcirc$), $LAM$ (green, $\square$) and $TT$ (blue, $\bigtriangleup$) and the RNA measures (dashed) $\mathcal{T}$ (magenta, $+$), $\mathcal{C}$ (cyan, $\bigtriangledown$) and $\mathcal{L}$ (black, $\diamond$) on the other hand, using windows of $256$ hours and mutual offset of $\Delta w=1$ hour. Dotted vertical lines indicate the ``location'' of the three embedding components $x(t)$, $x(t+\tau)$ and $x(t+2\tau)$, respectively.}
\label{fig:xcorr}
\end{figure}

Since we use embedding vectors taking information from different points in time into account, it is a relevant question which embedding component to consider as a reference when comparing the window-wise recurrence characteristics with the mean Dst amplitude. In our present setting, the two extreme solutions would be taking the first component $x(t)$ (i.e., taking information from present and \emph{future} Dst values) or the third component $x(t+2\tau)$ (present and \emph{past} values), whereas the second component $x(t+\tau)$ would provide some kind of trade-off. Indeed, Tab.~\ref{tab:correlations2} shows that the (negative) correlations with the Dst index are the strongest when taking the second component, i.e., considering a balanced information from both past and future of the observed process. This finding is further supported when considering the full cross-correlation functions between the mean Dst index and the six recurrence measures (Fig.~\ref{fig:xcorr}): all of them show their global minimum at values around $\tau$. Notably, the correlation coefficients when using $x(t)$ are of equal or even somewhat larger magnitude than those for $x(t+2\tau)$ except for $\mathcal{C}$ and $\mathcal{L}$, and the global minimum of the cross-correlation function is typically attained at values slighly below $\tau$ except for $\mathcal{C}$. This asymmetry suggests that information from the first embedding component might in fact be slighly more relevant than such from the third one, which could be related to a difference in the ``natural'' time-scales of the approaching of magnetic storms and the recovery phase. A further investigation and discussion of this result should be subject of future work.

\subsection{Estimated vs.\ baseline recurrence characteristics} \label{sec:bootstrapping}

The intervals obtained from the bootstrapping approaches (gray bars in Fig.~\ref{fig:results} -- corresponding to the upper and lower 5\% percentiles of the test distribution) -- display the behavior expected from our discussion in Section~\ref{sec:method}. Specifically, for all three RNA measures, the values within the individual windows are mostly above the upper 5\% quantiles of the distributions obtained by bootstrapping. This indicates that in comparison with the window-wise recurrence networks, bootstrapping state vectors from the entire time series introduces a considerable degree of randomness to the resulting networks, resulting in lower values of $\mathcal{T}$ and $\mathcal{C}$. This is because state vectors representing essentially different states of the system are mixed in the considered bootstrap samples. Since such states typically persist in time (cf.\, the time scales of several days at which the ACF in Fig.~\ref{fig:embedding}A decays), one has to expect a much higher degree of regularity in the networks obtained within running windows, expressed by higher values of both $\mathcal{T}$ and $\mathcal{C}$. In turn, if understanding the individual bootstrap samples as representatives of some stationary process that covers the full variety of dynamical patterns of Dst index variations, our results would be compatible with an almost constantly non-stationary (i.e., out-of-equilibrium) system, which is consistent with the modern view on the magnetosphere as described in the introduction of this manuscript. In this context, it is interesting to note that the distributions of RNA measures from the bootstrapped samples are considerably more narrow than the ranges of values obtained for the sliding windows, which is particularly well visible for $\mathcal{L}$.

Unlike for the RNA measures, the window-wise RQA characteristics scatter more symmetrically around the values obtained from the bootstrapped line length distributions, displaying both higher and lower values as time proceeds. Notably, positive ``anomalies'' exclusively arise during periods with strong magnetic storms, whereas values below the lower 5\% quantile of the line bootstrapping-based distributions are found only during periods without strong perturbations of the magnetosphere where the recorded Dst variations are dominated by short-term fluctuations. This general observation again underlines the fundamentally different conceptual foundations of the randomization procedures used for obtaining significance bounds for possible indications of regime changes in RQA and RNA.

\subsection{Dependence on methodological settings} \label{sec:settings}

\begin{figure*}
\centering
\resizebox{0.3\textwidth}{!}{\includegraphics*{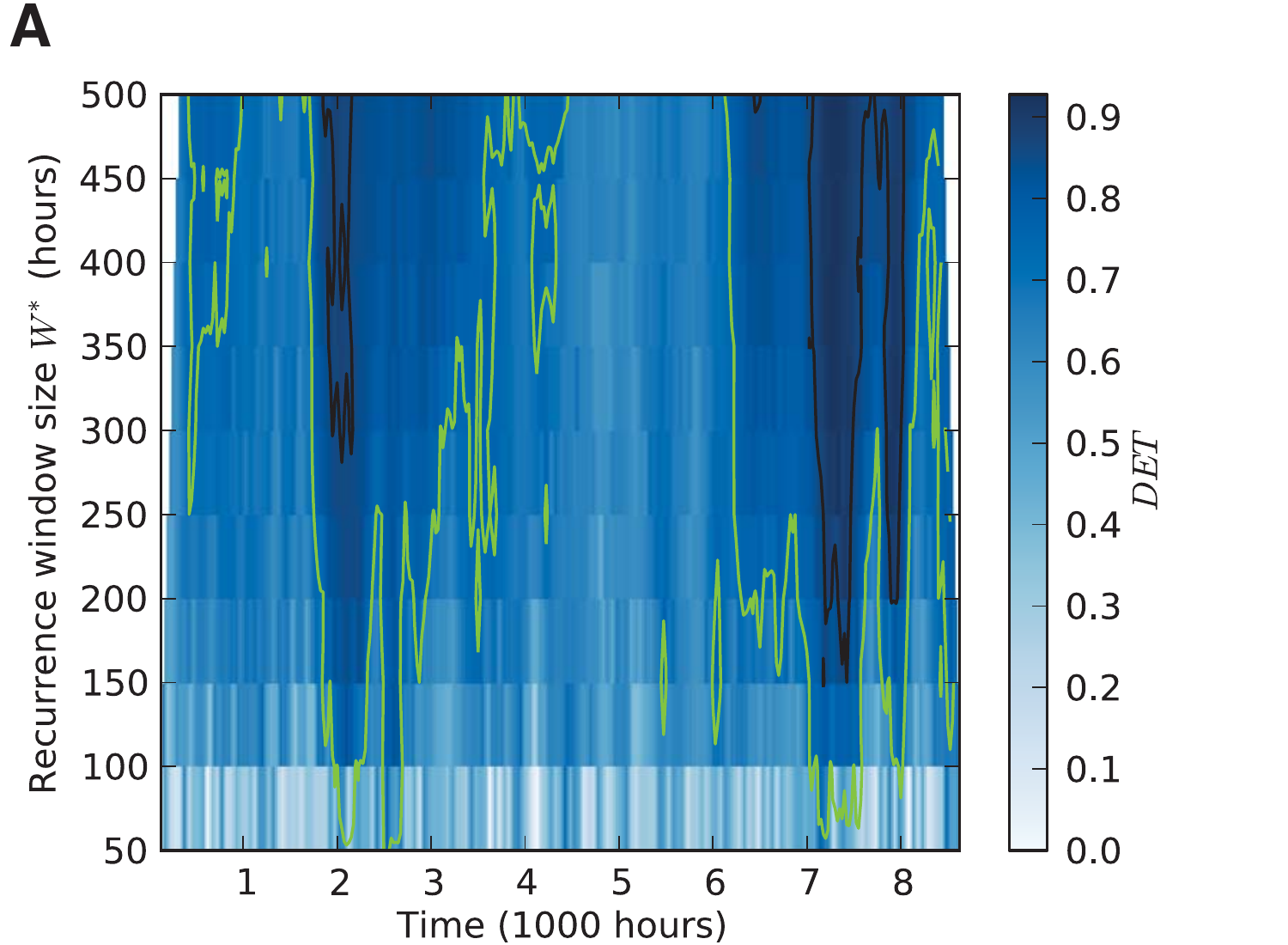}} \hfill
\resizebox{0.3\textwidth}{!}{\includegraphics*{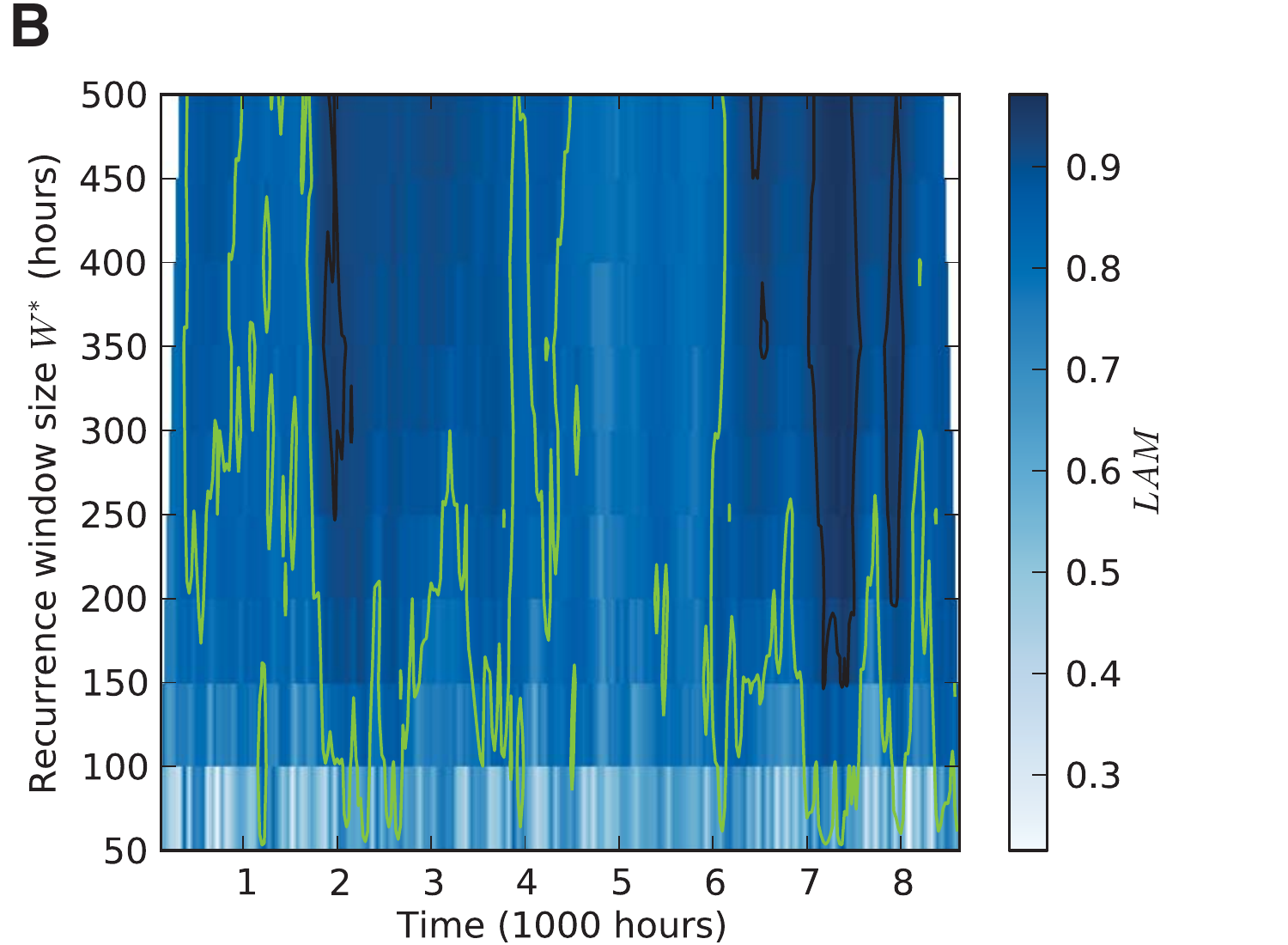}} \hfill
\resizebox{0.3\textwidth}{!}{\includegraphics*{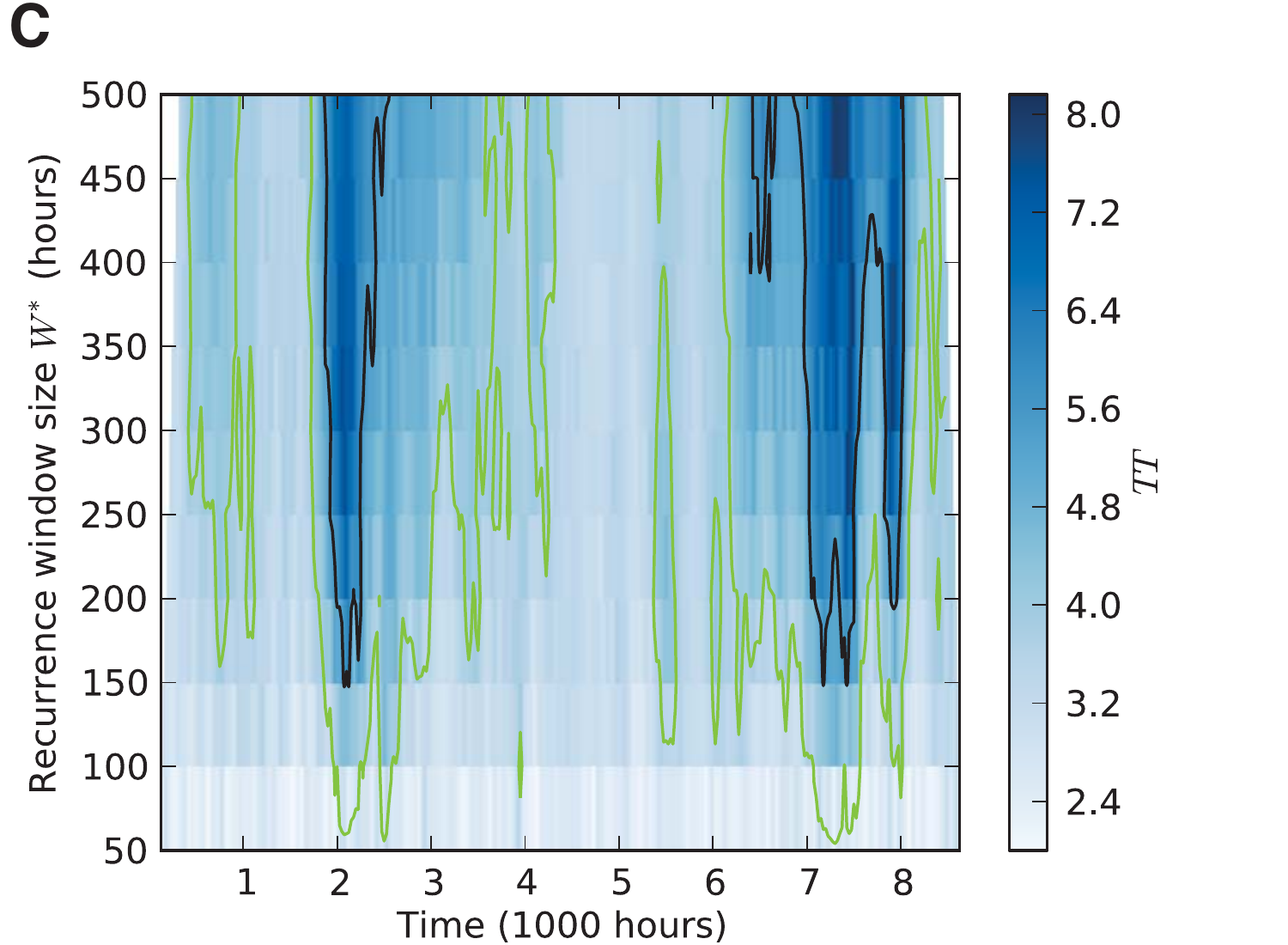}} \\
\resizebox{0.3\textwidth}{!}{\includegraphics*{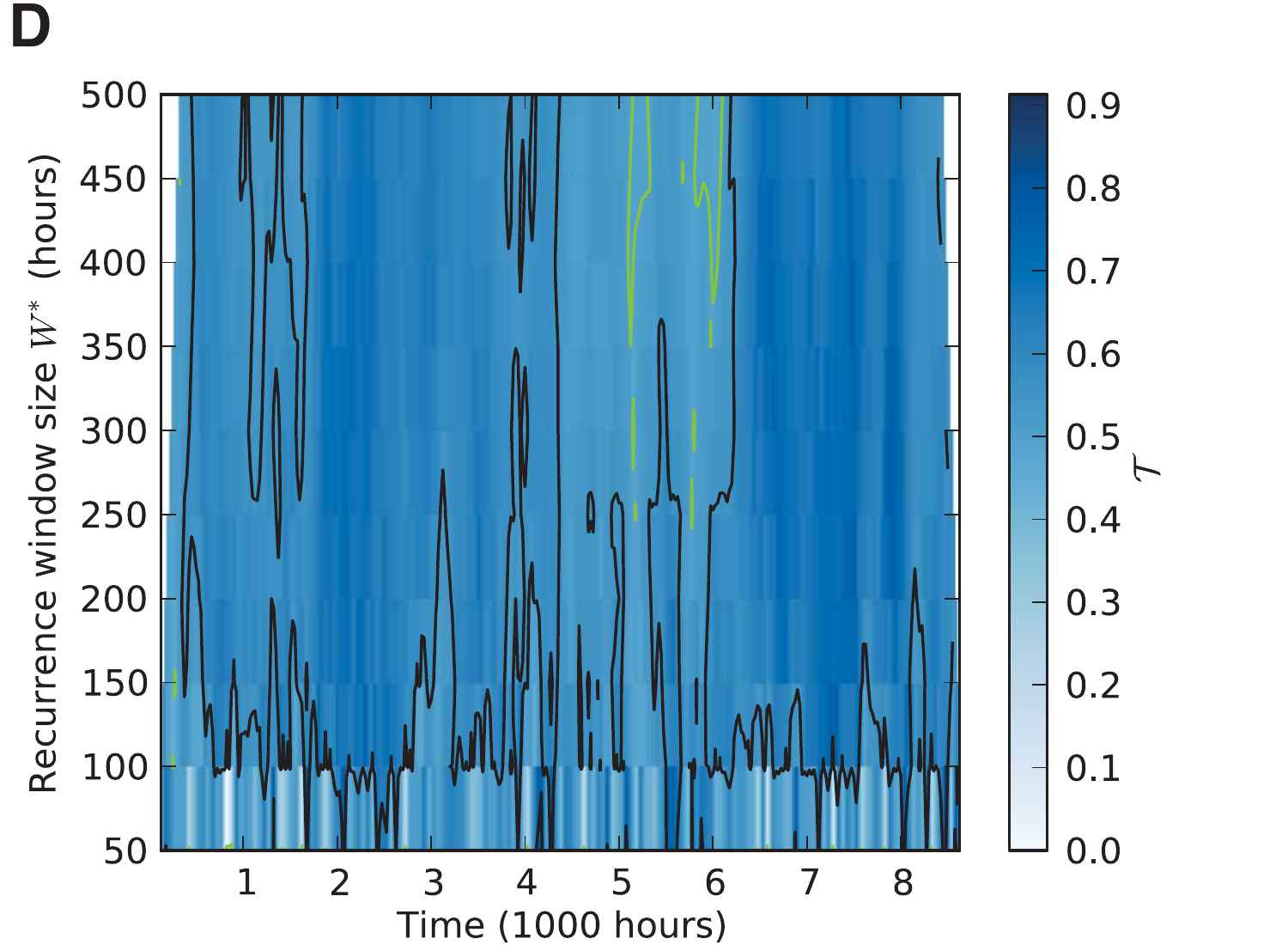}} \hfill
\resizebox{0.3\textwidth}{!}{\includegraphics*{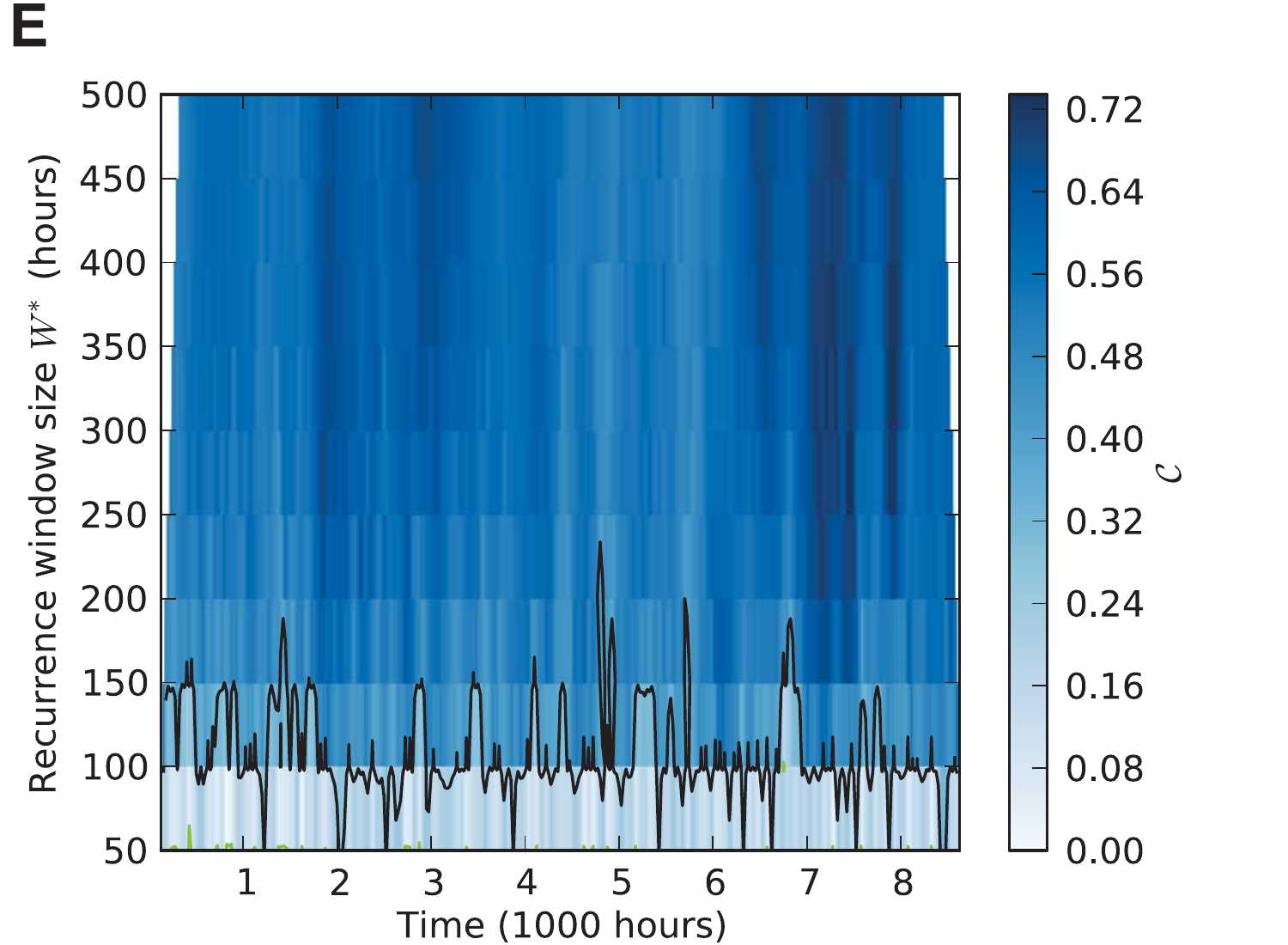}} \hfill
\resizebox{0.3\textwidth}{!}{\includegraphics*{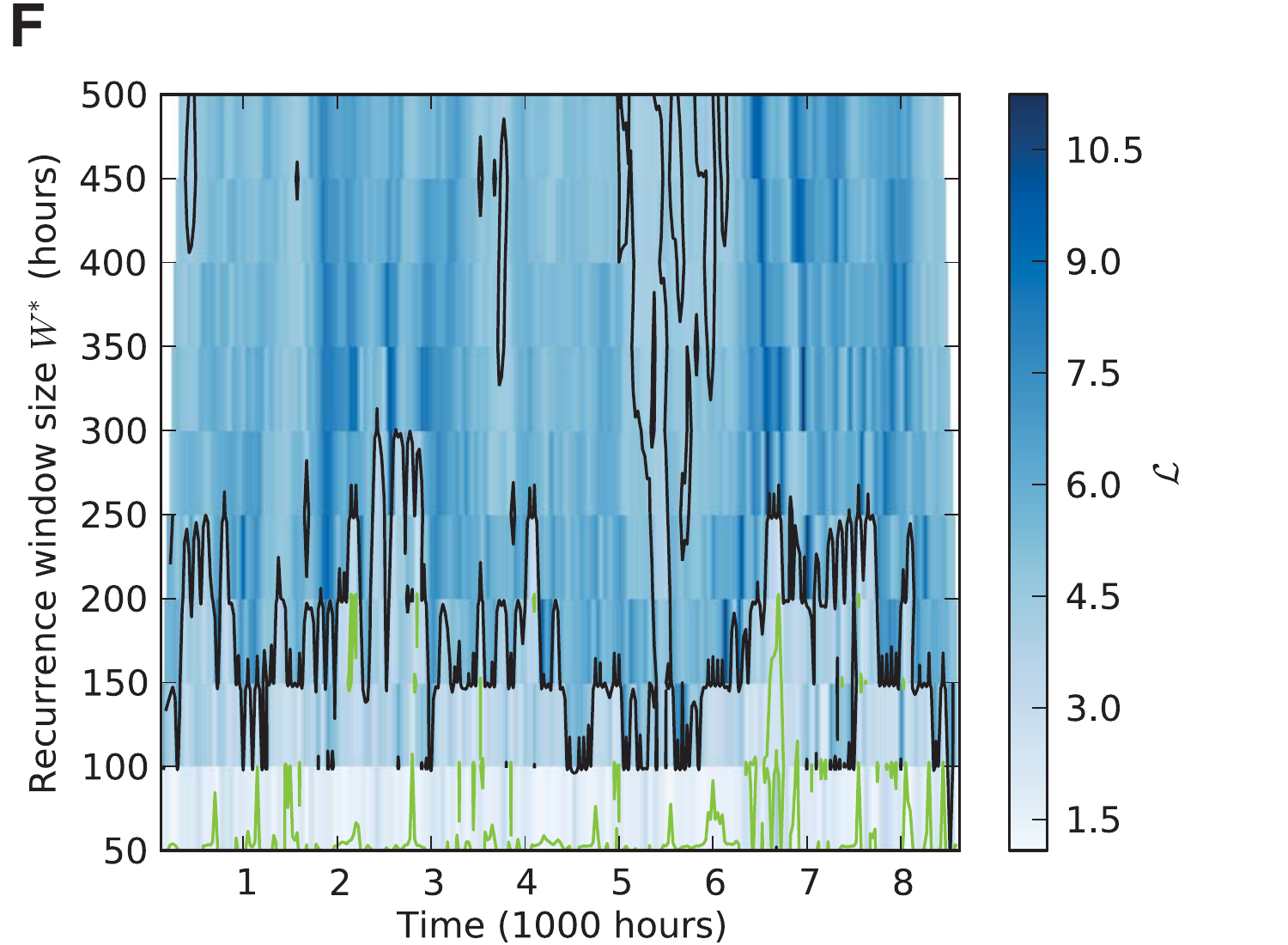}} \\
\caption{Dependence of the results of windowed RQA and RNA for all six parameters on the choice of the window size $w$. Colors indicate the value of the respective measure. Areas enclosed by black lines denote values above the respective upper 5\% quantiles of the bootstrapping based test distributions, whereas such enclosed by green lines correspond to values below the lower 5\% quantiles.}
\label{fig:robustness_window}
\end{figure*}

\begin{figure*}
\centering
\resizebox{0.3\textwidth}{!}{\includegraphics*{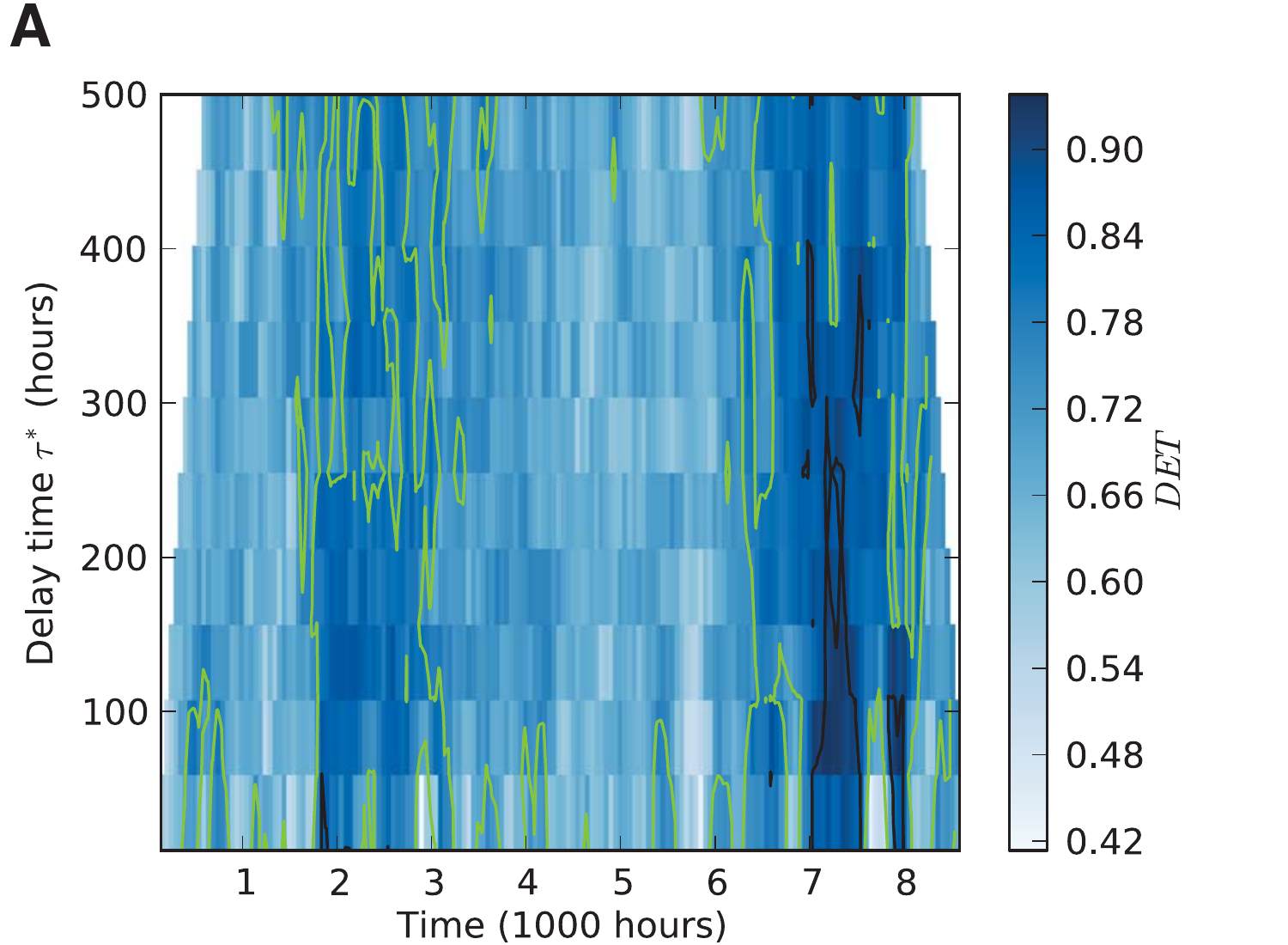}} \hfill
\resizebox{0.3\textwidth}{!}{\includegraphics*{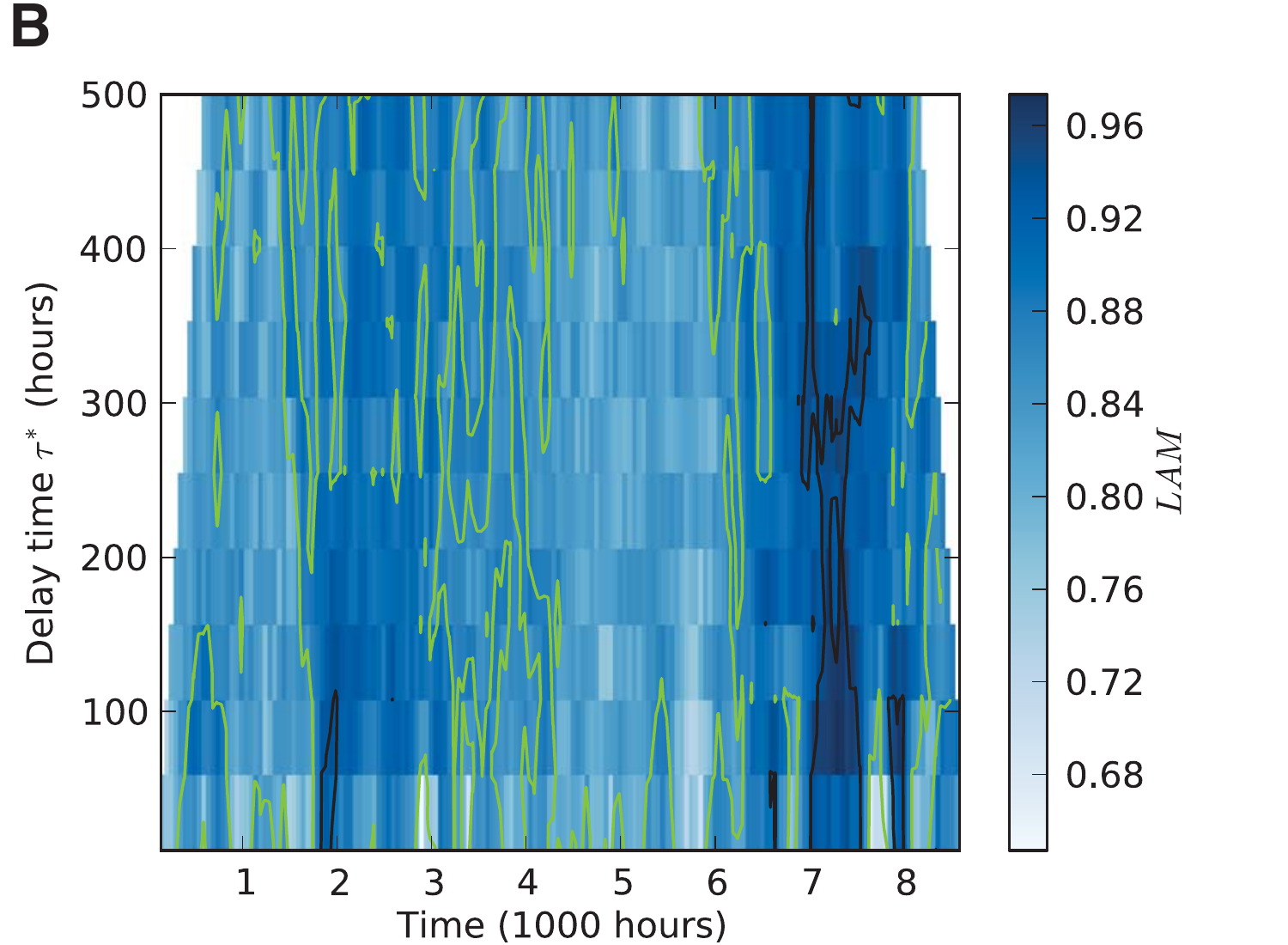}} \hfill
\resizebox{0.3\textwidth}{!}{\includegraphics*{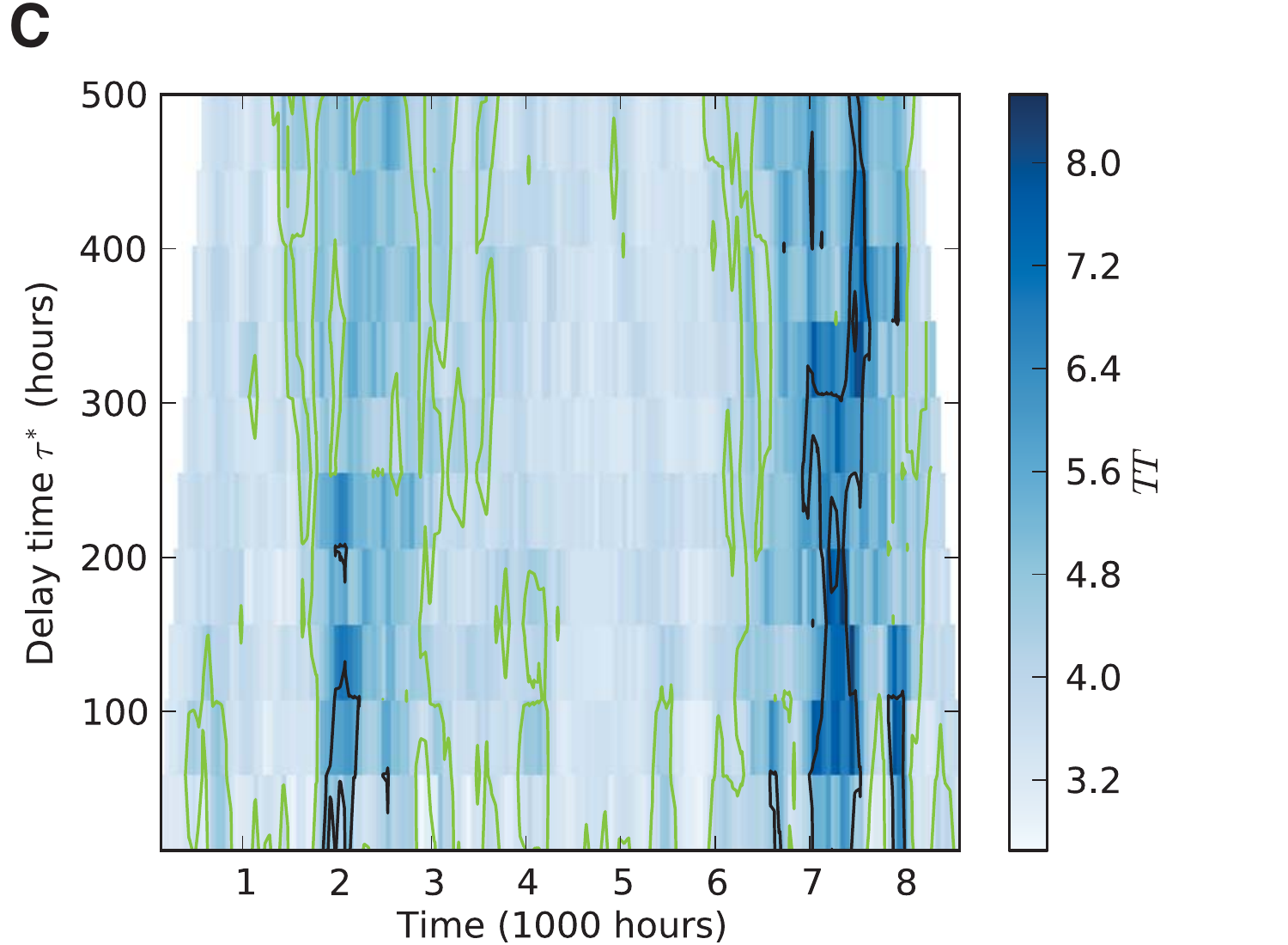}} \\
\resizebox{0.3\textwidth}{!}{\includegraphics*{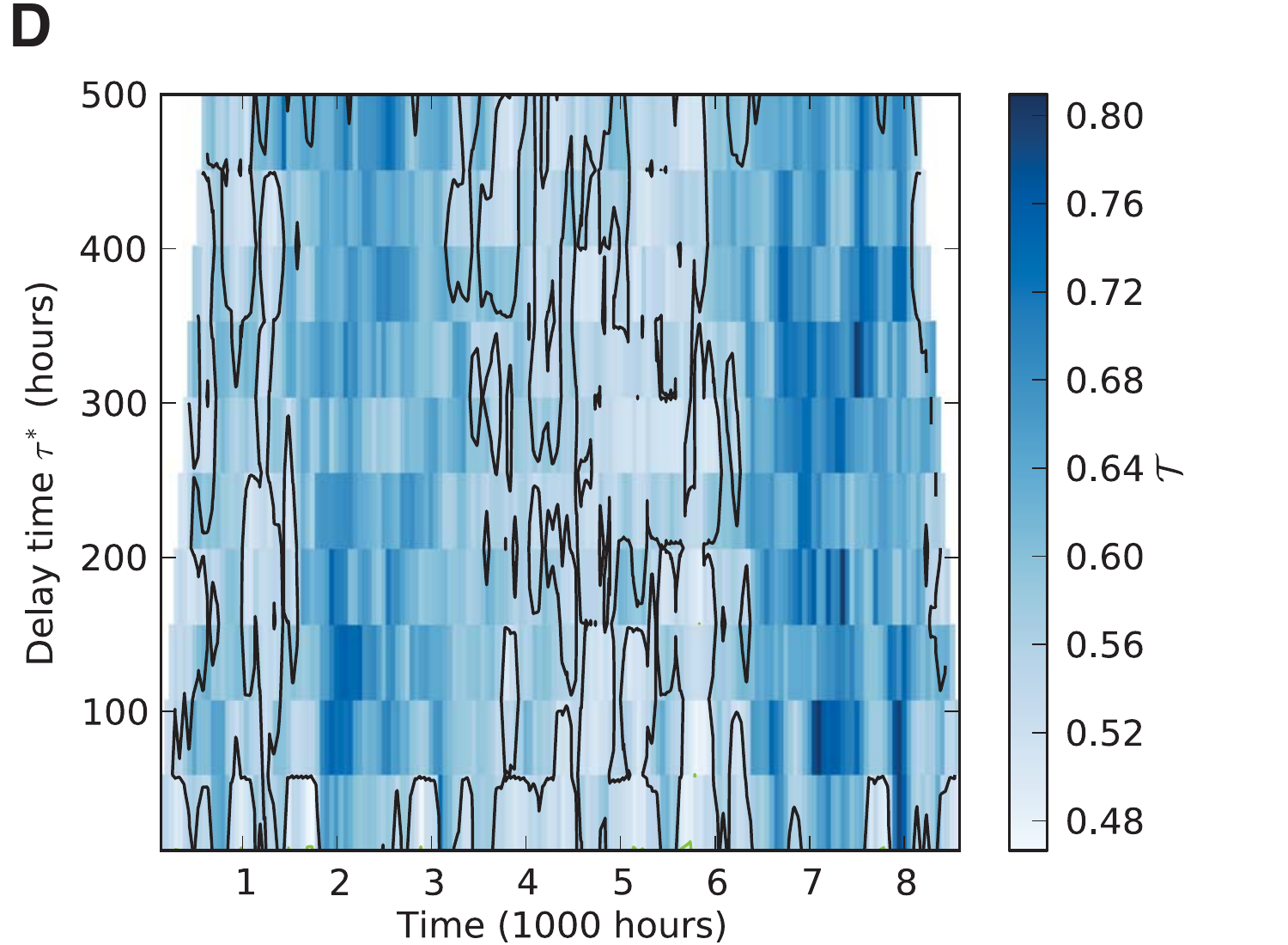}} \hfill
\resizebox{0.3\textwidth}{!}{\includegraphics*{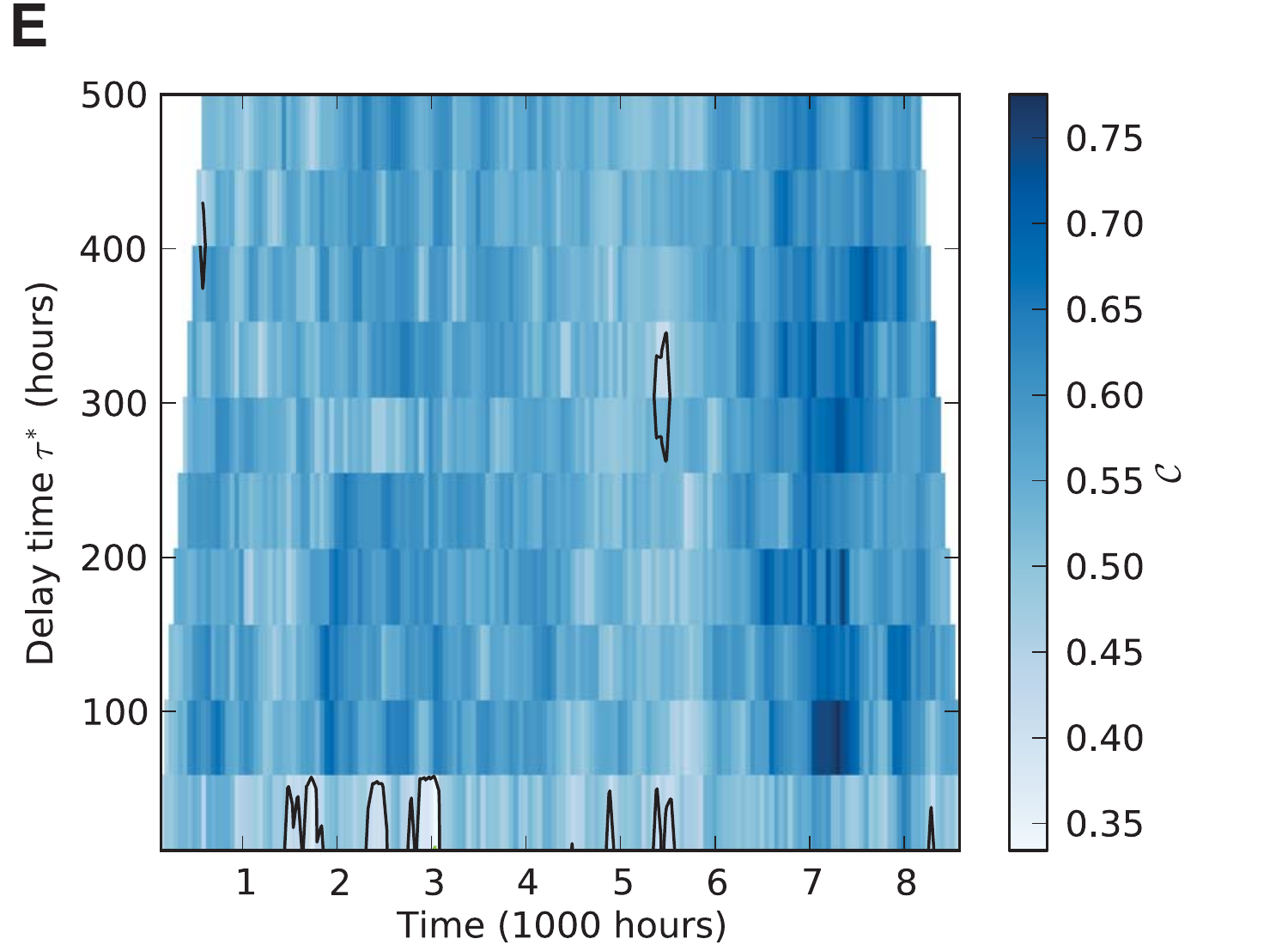}} \hfill
\resizebox{0.3\textwidth}{!}{\includegraphics*{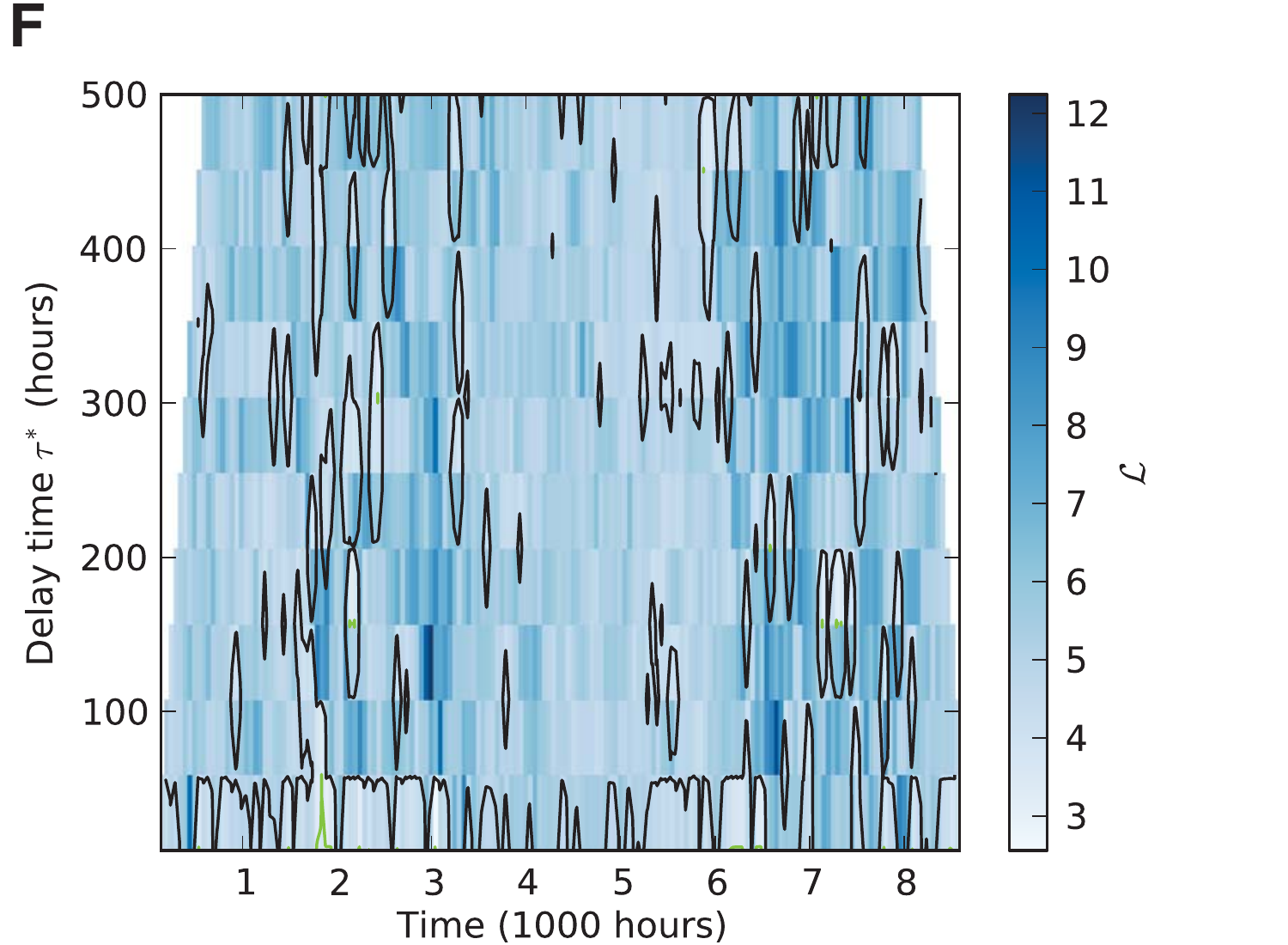}} \\
\caption{As in Fig.~\ref{fig:robustness_window} for the dependence on the embedding delay $\tau$. Note that for embedding delays $\tau$ much smaller than, say, 50 hours, the components of the embedding vectors still exhibit marked mutual interdependencies (see Fig.~\ref{fig:embedding}A), which may result in systematic biases of the computed recurrence characteristics.}
\label{fig:robustness_delay}
\end{figure*}

The recurrence framework used in this study includes a set of parameters (especially embedding dimension $m$ and delay $\tau$, recurrence rate $RR$ and window size $w$) that need to be carefully chosen. Frequently, this selection is not unique, but based on some educated guesses or rules-of-thumb. Typically, the results of RQA and RNA depend only moderately on the exact choice of these parameters, i.e., they do not change qualitatively if the setting is changed within some (often a priori unknown) acceptable limits~\cite{Donges2011}. In order to study if this presumed robustness also applies to our results discussed above, Figs.~\ref{fig:robustness_window} and \ref{fig:robustness_delay} display the six recurrence measures for different values of the window length $w$ and embedding delay $\tau$, respectively. We find that variations of both parameters indeed do not change the temporal variability of all RQA and RNA measures markedly, demonstrating that the obtained results are qualitatively robust under changes of the intrinsic parameters of our methods. Note that according to our ``confidence'' intervals from the bootstrapping-based approaches discussed above, we display the contour lines indicating values outside the respective bounds in Figs.~\ref{fig:robustness_window} and \ref{fig:robustness_delay} only for reference.

When varying the recurrence rate $RR$, the RQA measures are known to commonly change only weakly in some gradual way. Among the RNA measures, we can expect that $\mathcal{T}$ and $\mathcal{C}$ take similar values over a reasonable range of recurrence rates (see \cite{Donges2012} for some discussion on the operational window), whereas $\mathcal{L}$ changes with inverse proportionality to the recurrence threshold $\varepsilon$ \cite{Donner2010,Donges2012}. The latter theoretical finding provides an important argument for fixing the recurrence rate rather than $\varepsilon$ in running window-based recurrence analyses. 

The other parameter whose influence is not explicitly studied here is the embedding dimension $m$. However, since this dimension can only take very few discrete values (balancing complexity versus data availability), there are not many possibilities for selecting values different from $m=3$ as chosen here. This choice is motivated by a trade-off between different considerations: (i) the necessity of embedding to appropriately represent the observed dynamics, (ii) the finding that the ``optimal'' embedding dimension for the complete time series might be much larger (i.e., at least 5), (iii) the requirement that the embedding vectors should not cover much longer time intervals than the window size within which the recurrence properties are evaluated, and (iv) the fact that embedding can induce spurious structures even for uncorrelated random processes \cite{Thiel2006}. Taking all these aspects together, $m=3$ is a reasonable choice, and further detailed discussion of the impact of other values will most probably not provide any fundamental new insights to the topic of this work.

\section{Discrimination between storm and non-storm periods}\label{sec:discrimination}

In the following, we investigate in some detail how well the values of the different recurrence characteristics discriminate between storm and non-storm periods. For this purpose, we employ the heuristic classification already used in Section~\ref{sec:data}. Specifically, for running windows of width $w=256$ hours and mutual offset $\Delta w=1$ hour, we consider the windows with indices 1501 to 3150 and 6151 to 8000 (i.e., with midpoints between 8 March and 16 May 2001 and 17 September and 4 December, respectively) as storm periods, the others as non-storm periods.

\subsection{Recurrence characteristics}

\begin{figure}
\centering
\resizebox{0.485\textwidth}{!}{\includegraphics*{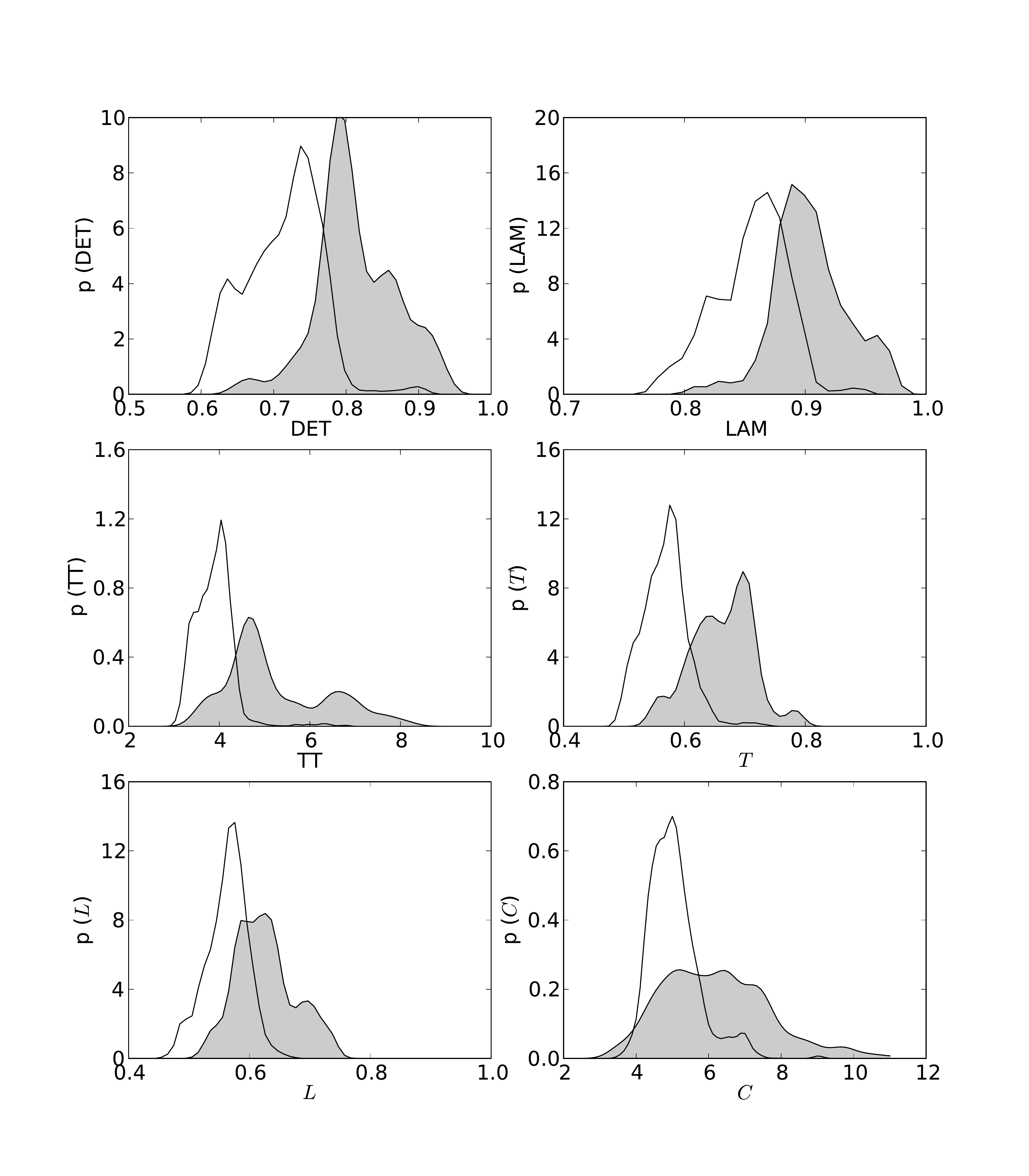}}
\caption{Probability density functions (Gaussian kernel estimates) of the different RQA and RNA measures (obtained for running windows of width $w=256$ hours and mutual offset $\Delta w=1$ hour) during storm (gray area) and non-storm (white area) periods (heuristic classification as detailed in text).}
\label{fig:pdf}
\end{figure}

Figure~\ref{fig:pdf} presents the PDFs of all six characteristics estimated separately for both types of conditions. As for the hourly Dst index itself (Fig.~\ref{fig:pdf_dst}), we observe considerable overlaps between the empirical distributions for both cases, but the mean values appear clearly separated given the respective variances within the two subsamples. In order to characterize the significance of the mutual differences between the distributions for storm and non-storm periods, a variety of statistical approaches could be used (see \cite{Zou2010} for a corresponding discussion and one example regarding RQA and RNA for discriminating between intertwined periodic and chaotic windows in some paradigmatic model system). Regarding a classical one-way ANOVA (analysis of variance, quantifying the difference between the mean values of two samples given their respective variances) approach \cite{Barlow1989}, all six recurrence measures clearly do not pass the $F$-test for equality of the means and the Wilcoxon rank-sum test \cite{Wilcoxon1945} for equality of the medians at very high confidence levels, indicating that the distributions of all RQA and RNA measures for storm and non-storm periods are significantly different.

One particularly instructive approach for quantitatively studying the discriminative skills of statistical methods based on a given data classification is the so-called receiver-operating characteristics (ROC) analysis \cite{Fawcett2006}. Here, a variable threshold is applied to each of the six measures, and for each threshold value, the classification of values (above/below the threshold) is compared with the given classification of the data into storm and non-storm periods. The rates of true and false ``positive'' classifications (TPR and FPR, respectively) of a storm period based on each RQA or RNA measure are continuously monitored and provide a characteristic closed curve (the ROC curve) in the (FPR,TPR) plane. 

\begin{figure*}
\centering
\resizebox{0.485\textwidth}{!}{\includegraphics*{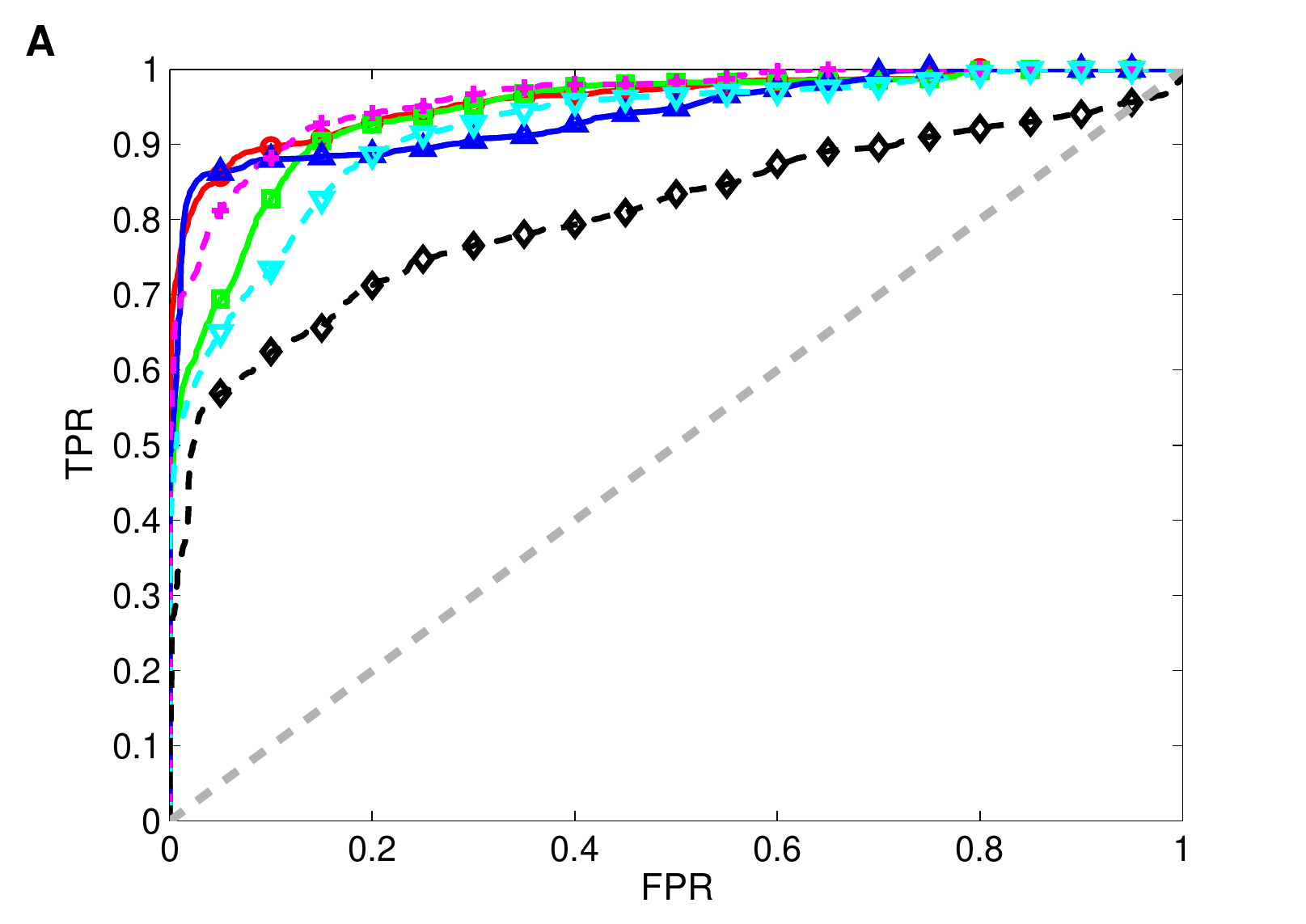}} \hfill
\resizebox{0.485\textwidth}{!}{\includegraphics*{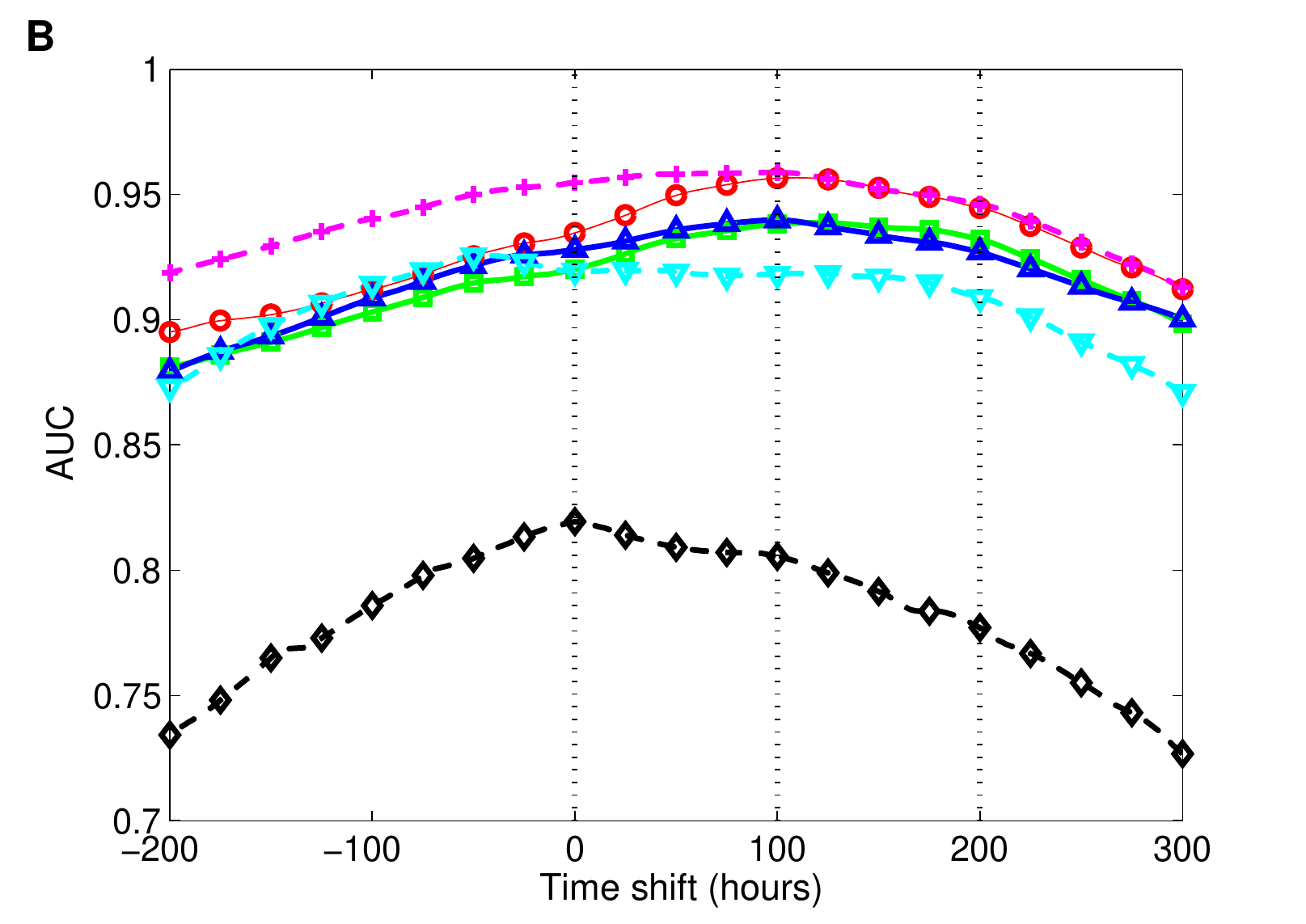}}
\caption{Results of receiver-operating characteristics (ROC) analysis for the RQA and RNA measures (line colors and symbols as in Fig.~\ref{fig:xcorr}). (A) ROC curves using windows of $256$ hours and mutual offset of $\Delta w=1$ hour, given the heuristic splitting of the Dst time series into five storm and non-storm periods (see text) as a reference. Following the results from Fig.~\ref{fig:xcorr}, we use the second component $x(t+\tau)$ of the embedding vector as a reference for relating the windows considered for Dst and the recurrence characteristics. The dashed diagonal line indicates the expected performance of a random discrimination (TPR$=$FPR). (B) Dependence of the area under the ROC curve (AUC) on the mutual offset between the Dst windows and the windows for which the recurrence measures are considered. Here, a time shift of zero corresponds to the timing of each Dst value coinciding with that of the first component of the embedding vector. Dotted vertical lines indicate the ``location'' of the three embedding components $x(t)$, $x(t+\tau)$ and $x(t+2\tau)$, respectively.}
\label{fig:roc_fixed}
\end{figure*}

Figure~\ref{fig:roc_fixed}A shows the ROC curves of all six measures. The closer the ROC curve lies at the upper left corner or, more precisely, the stronger it deviates from the main diagonal TPR$=$FPR corresponding to a random ``prediction'', the better are the skills of the respective measure to discriminate between both types of (longer-term) magnetospheric ``states''. Consequently, this skill can be measured in terms of the area under the ROC curve, AUC. 

Figure~\ref{fig:roc_fixed}B displays the dependence of AUC on the choice of the reference time for comparing windowed Dst and RQA/RNA measures, which can be interpreted in a similar way as the cross-correlation functions in Fig.~\ref{fig:xcorr}. We find that maxima of AUC indicating the best discriminatory skills of the recurrence measures are mostly concentrated around the timing associated with the second embedding component. Notable exceptions are $\mathcal{L}$ and $\mathcal{C}$, which however show the worst performance with respect to the criterion considered here. In general, given the coarse-grained classification between storm and non-storm periods as discussed above, $\mathcal{T}$ generally provides the highest AUC values, followed by $DET$, $TT$ and $LAM$, all of which still provide a very good discrimination. To a certain degree, the AUC values of $\mathcal{C}$ also suggest a reasonable distinction between storm and non-storm periods, whereas the results for $\mathcal{L}$ are not satisfactory in this respect.

\begin{table}
\centering
\begin{tabular}{|c|ccc|}
\hline
& AUC($0$) & AUC($\tau$) & AUC($2\tau$) \\
\hline
$DET$ & 0.935 & 0.956 & 0.945 \\
$LAM$ & 0.920 & 0.938 & 0.932 \\
$TT$ & 0.928 & 0.940 & 0.927 \\
$\mathcal{T}$ & 0.955 & 0.959 & 0.946 \\
$\mathcal{C}$ & 0.919 & 0.918 & 0.909 \\ 
$\mathcal{L}$ & 0.819 & 0.806 & 0.777 \\
\hline
\end{tabular}
\vspace{0.2cm}
\caption{Area under the ROC curve (AUC) indicating the performance of the six recurrence measures as discriminators between the (heuristically classified) storm and non-storm periods. For defining the running windows for the recurrence-based measures, the first, second and third components of the embedding vectors ($x(t)$, $x(t+\tau)$ and $x(t+2\tau)$, respectively) have been considered.}
\label{tab:auc_fixed}
\end{table}

Notably, the ranking of the six recurrence measures according to AUC (cf.~also Tab.~\ref{tab:auc_fixed}) differs somewhat from that based upon the correlation analysis in Section~\ref{sec:performance}. Specifically, $\mathcal{T}$ still performs best among all characteristics, closely followed by $DET$, whereas $TT$ provides lower AUC values than expected. This discrepancy can be explained by the fact that we have considered here a rather coarse-grained classification between storm and non-storm periods, whereas correlations directly rely on the explicit window-wise mean Dst values, which can be the same for a considerable number of windows as illustrated in Fig.~\ref{fig:pdf}. We will return to this aspect in Section~\ref{sec:adaptive}.

\subsection{Comparison with other methods}

In previous works, changes in the dynamical properties of the Dst index have been extensively studied and related to different levels of organization of magnetospheric dynamics during storms and non-storm periods. In the following, we will briefly review the main corresponding findings and compare them with the results of our recurrence-based analysis.

In~\cite{Balasis2006,Balasis2011b}, temporal changes in the persistence of the Dst data have been studied in terms of the associated Hurst exponent $H$ estimated by some fractal wavelet spectral approach and rescaled-range analysis, respectively. Both studies reported a marked increase in the estimated exponents during storm periods, pointing to some large-scale and long-term organization of magnetospheric fluctuations. The corresponding results were qualitatively confirmed in a recent study \cite{Donner2013} utilizing a simple complexity measure (the linear variance decay (LVD) dimension density $\delta_{LVD}$) based on the auto-correlation function of the data under study.

References~\cite{Balasis2008,Balasis2009,Balasis2011a} applied the classical concept of Shannon entropy $S$ in conjunction with Tsallis' non-extensive entropy $S_q$ as well as some other entropy measures in order to characterize the degree of dynamical disorder. Their results consistently revealed a marked decrease of entropic  characteristics during magnetic storms, pointing again towards an elevated degree of dynamical organization and associated determinism in the magnetosphere.

In two recent papers~\cite{Balasis2009,Balasis2013}, a suite of complementary entropy concepts have been utilized, which are not solely rooted in information theory and statistical mechanics (like Shannon or Tsallis entropy), but make use of distances among vectors in the system's reconstructed phase space. Specifically, the approximate entropy (AppEnt) \cite{Pincus1991}, sample entropy (SampEnt) \cite{Richman2000} and fuzzy entropy (FuzzyEnt) \cite{Chen2007} used in the aforementioned works are based on a conceptual foundation closely related to that of the recurrence characteristics employed in this work (see \cite{Balasis2013} for more details on the respective methods and their relation to recurrence analysis). In this spirit, when using these measures we may expect results that are generally comparable with those of recurrence analysis as discussed above. 

In the following, we provide an in-depth analysis on how well different previously considered dynamical characteristics discriminate between the state of the magnetosphere during storm and non-storm periods based on the Dst index data. For this purpose, we apply the methods listed below with the following parameters:
\begin{itemize}
\item The Hurst exponent $H$ is estimated from the slopes of the wavelet power spectral densities in the frequency range between 2 and 128 hours \cite{Balasis2006}.
\item The LVD dimension density $\delta_{LVD}$ \cite{Donner2013} is computed using $N=100$ embedding components mutually shifted by $\tau=1$ hour and a threshold for the fraction $f$ of explained variance at $f=0.95$.
\item The Shannon entropy $S$ is calculated for a binary partition (with the mean value as threshold) of the underlying data. The block ($H_n$) and Tsallis entropies ($S_q)$ are computed for the same partition based on a block (word) length of $n=2$ (for details, see \cite{Balasis2009}). The value of the Tsallis $q$ index utilized for the calculation of non-extensive Tsallis entropy $S_q(q)$ is selected to be 1.8, as indicated in \cite{Balasis2008} and more recently in \cite{Balasis2011c}.
\item The Kolmogorov entropy $h$ is estimated from the linear scaling behavior of the block entropy per symbol, $H_n/n$ for larger $n$.
\item The T-complexity $T$ is calculated using $n=2$ and $m=1$, respectively (for details, see \cite{Balasis2009}).
\item The approximate, sample and fuzzy entropies are computed using embedding dimension $m=2$, $\tau=1$ hour and a distance threshold $\varepsilon=0.65 \sigma$ (with $\sigma$ being the standard deviation of the data within the considered time window) to comply with the setting of previous work~\cite{Balasis2013}. For FuzzyEnt, the fuzzy membership function described in~\cite{Balasis2013} is utilized with $n=2$.
\end{itemize}
\noindent

\begin{figure*}
\centering
\resizebox{0.85\textwidth}{!}{\includegraphics*{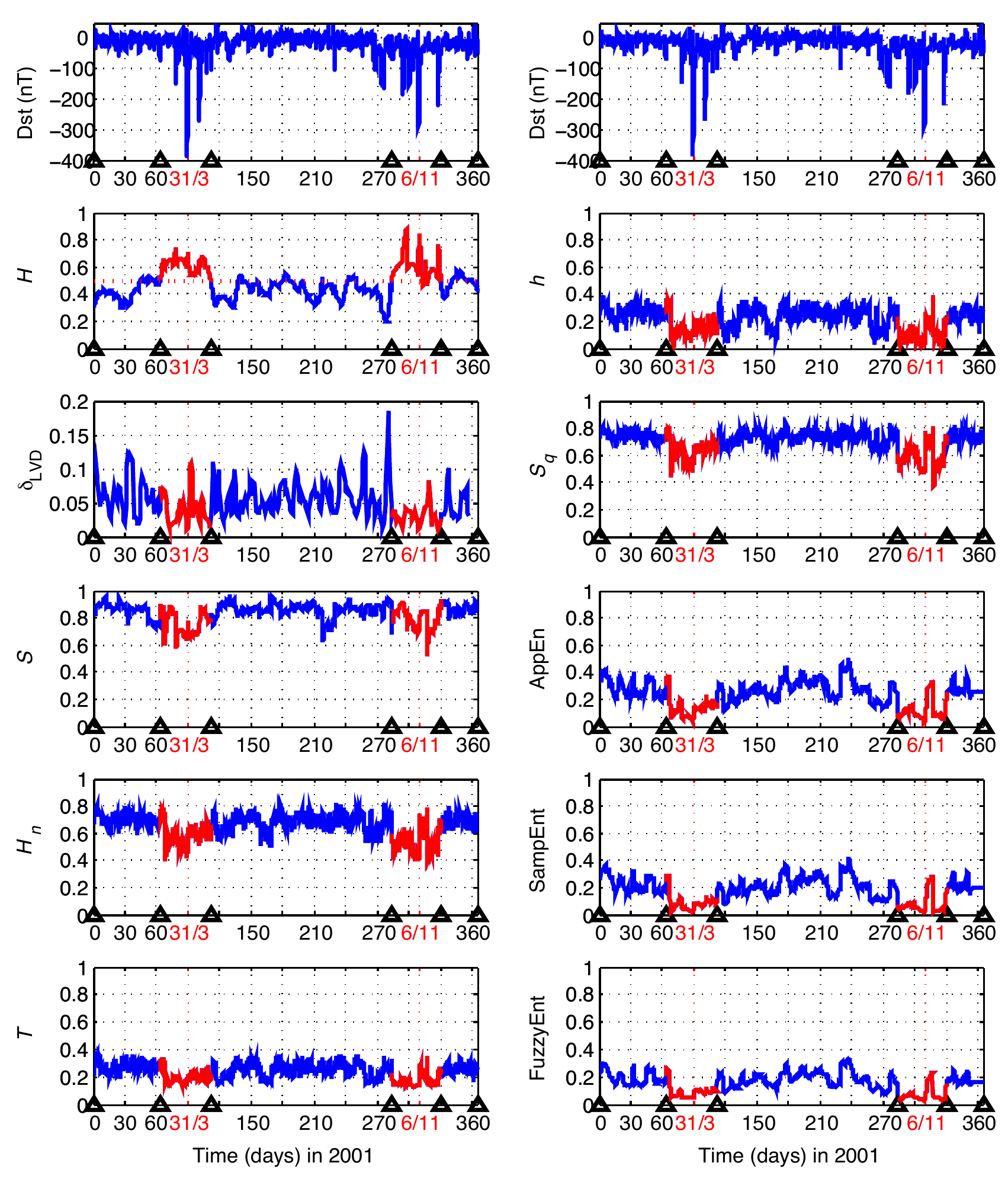}}
\caption{Time evolution of Dst index (top panels) and its entropy and correlation-based characteristics: Hurst exponent $H$, LVD dimension density $\delta_{LVD}$, Shannon entropy $S$, block entropy $H_n$, T-complexity $T$ (left panels, from top to bottom), Kolmogorov entropy $h$, Tsallis entropy $S_q$, AppEnt, SampEnt and FuzzyEnt (right panels, from top to bottom). Further details on the respective methodological settings are provided in the text. Red colors mark the storm periods II and IV from Fig.~\ref{fig:data}. Dotted red vertical lines indicate the timing of the two major storms during the considered interval of observations. For the Hurst exponent $H$, the dashed horizontal line gives the value of $0.5$ discriminating between persistent and anti-persistent dynamics.}
\label{fig:others}
\end{figure*}

For brevity, we present only the results for a window width of $w=256$ hours and a mutual offset of $\Delta w=1$ hour. A summary of the temporal variations of all measures is provided in Fig.~\ref{fig:others}. Note that as for the recurrence measures (Section~\ref{sec:performance}) and the LVD dimension density \cite{Donner2013}, the performance of all characteristics in discriminating between storm and non-storm periods depends on the window width as well as the definition of the storm period. However, we will not discuss this aspect further in the present work.

\begin{figure*}
\centering
\resizebox{0.97\textwidth}{!}{\includegraphics*{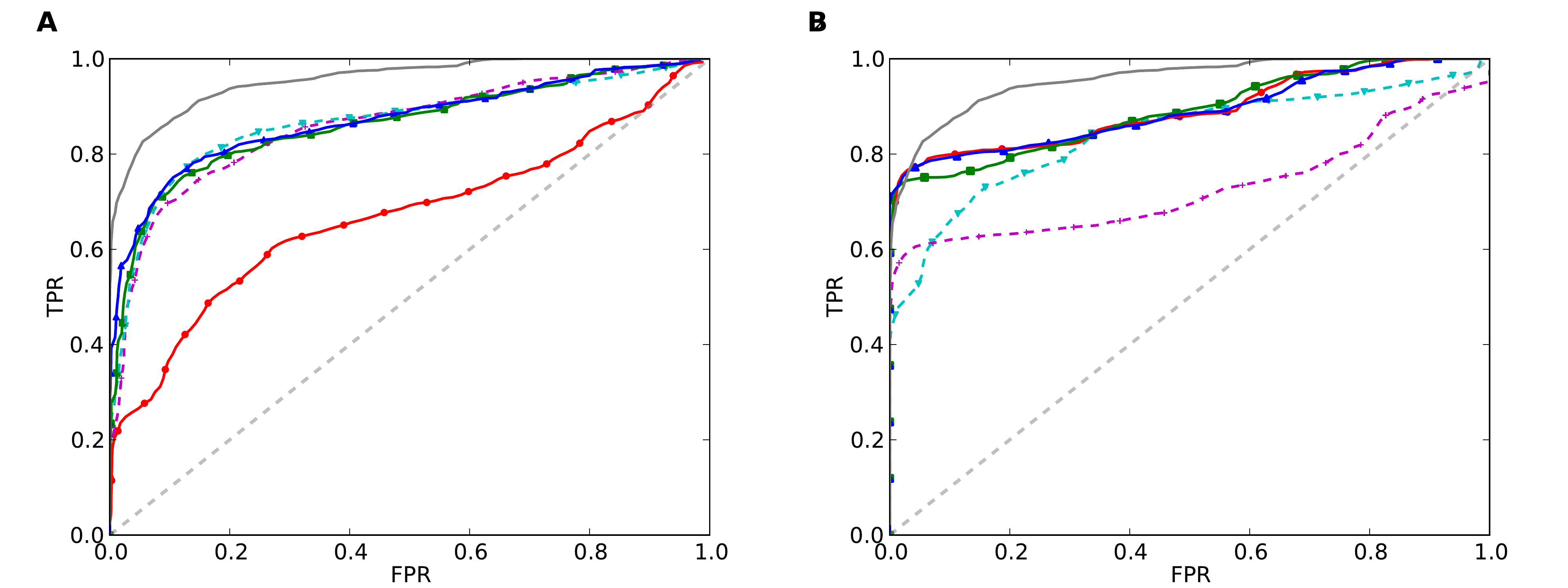}}
\caption{ROC curves for other measures recently considered in the literature ($w=256$ hours, $\Delta w=1$ hour). (A) Symbolic dynamics based characteristics Shannon entropy $S$ (red, $\bigcirc$, solid, AUC=0.670), block entropy $H_n$ (green, $\square$, solid, AUC=0.862), Tsallis entropy $S_q$ (blue, $\bigtriangleup$, solid, AUC=0.871), T-complexity $T$ (magenta, $+$, dashed, AUC=0.860) and Kolmogorov entropy $h$ (cyan, $\bigtriangledown$, dashed, AUC=0.867). (B) Phase space-based entropic quantities AppEnt (red, $\bigcirc$, solid, AUC=0.893), SampEnt (green, $\square$, solid, AUC=0.890) and FuzzyEnt (blue, $\bigtriangleup$, solid, AUC=0.892) together with the correlation-based measures Hurst exponent $H$ (magenta, $+$, dashed, AUC=0.730) and LVD dimension density $\delta_{LVD}$ (cyan, $\bigtriangledown$, dashed, AUC=0.845). For reference, the corresponding ROC curve for the recurrence network transitivity $\mathcal{T}$ is displayed as well (solid gray line).}
\label{fig:roc_others_fixed}
\end{figure*}

Our results shown in Fig.~\ref{fig:roc_others_fixed} reveal that the dynamic entropies (block, Tsallis and Kolmogorov entropy, T-complexity as well as AppEnt, SampEnt and FuzzyEnt) and the LVD dimension density provide a reasonable discrimination with AUC values of 0.84 and higher, whereas the purely statistical Shannon entropy and the Hurst exponent yield significantly poorer performance. The three phase space-based entropies yield the highest AUC values among all measures, which are, however, still significantly lower than those for the four best-suited recurrence characteristics. One possible reason for this is that the parameters of the recurrence analysis (especially the embedding parameters) have been adjusted to the specific data set, whereas the ``embedding'' dimension and delay have been chosen at the lowest possible values for the entropy estimations. Notably, the latter setting might not allow resolving all essential time-scales of variations associated with magnetospheric dynamics (unlike for the choice embedding parameters used for RQA and RNA analysis). By further tuning the parameters for AppEnt, SampEnt and FuzzyEnt, a systematic improvement of the skills of these methods could be achieved. As a somewhat surprising result, we emphasize that the correlation-based LVD dimension density performs not much worse than the dynamic entropies (even though it is based on some specific statistical model of the correlations that does not necessarily provide a reasonable approximation of the actual data properties). Reversing the aforementioned argument regarding the phase space-based entropies, this finding could be related to the relatively large number of embedding components used in estimating this characteristic.

\section{Discrimination using a data-adaptive classification}\label{sec:adaptive}

Instead of the previous fixed and thus rather inflexible classification of storm and non-storm periods, in the following we turn to some more data adaptive strategy by defining for each running window a storm as being present if the mean Dst value taken over the window is below some threshold value $\left<\mbox{Dst}\right>^*$. In the following, we will commonly consider $\left<\mbox{Dst}\right>^*=-30$~nT unless stated otherwise. The choice of this threshold value will be justified in the course of the analysis described in this section.

\subsection{Recurrence based characteristics}

\begin{figure*}
\centering
\resizebox{0.485\textwidth}{!}{\includegraphics*{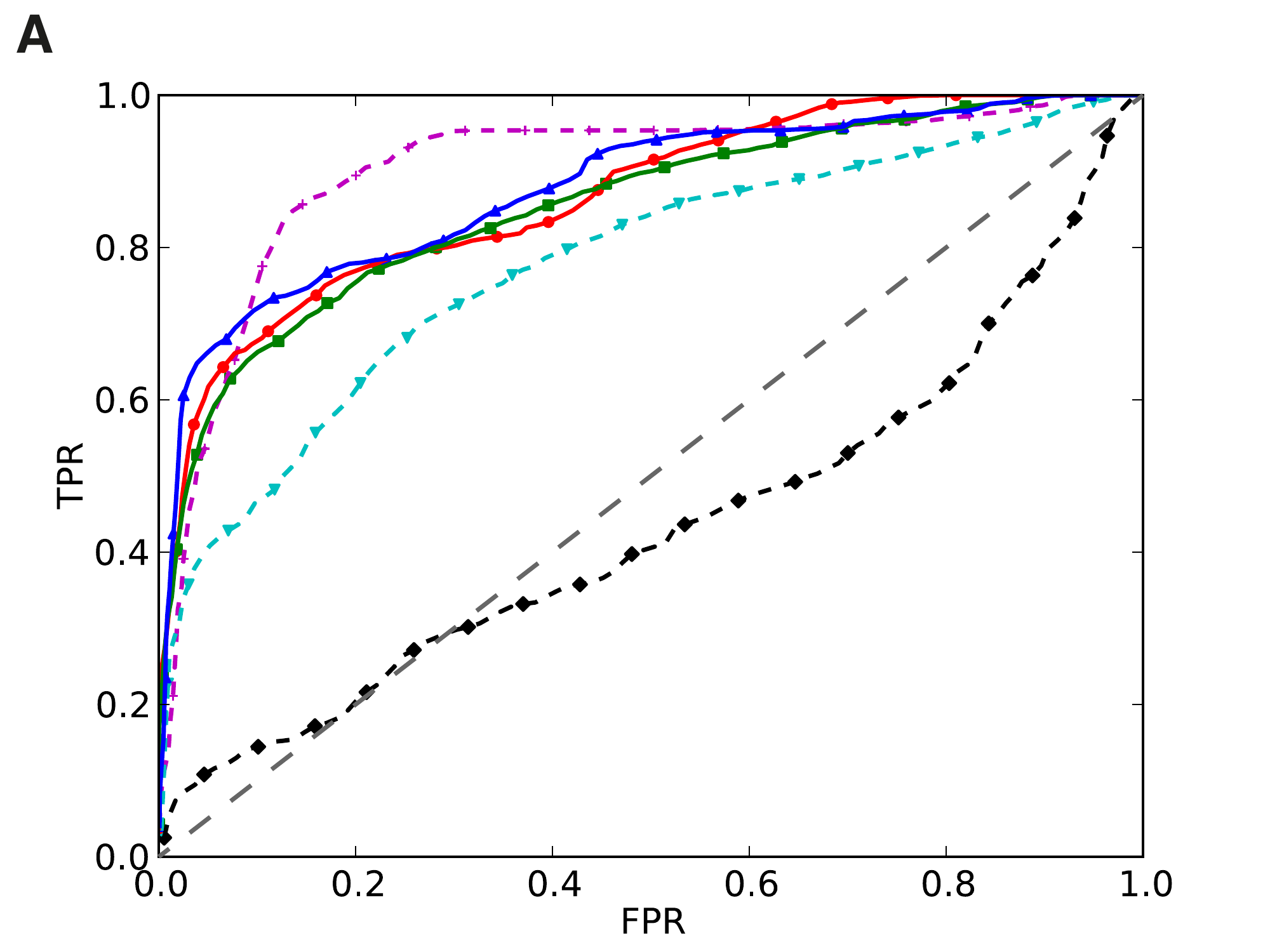}} \hfill
\resizebox{0.485\textwidth}{!}{\includegraphics*{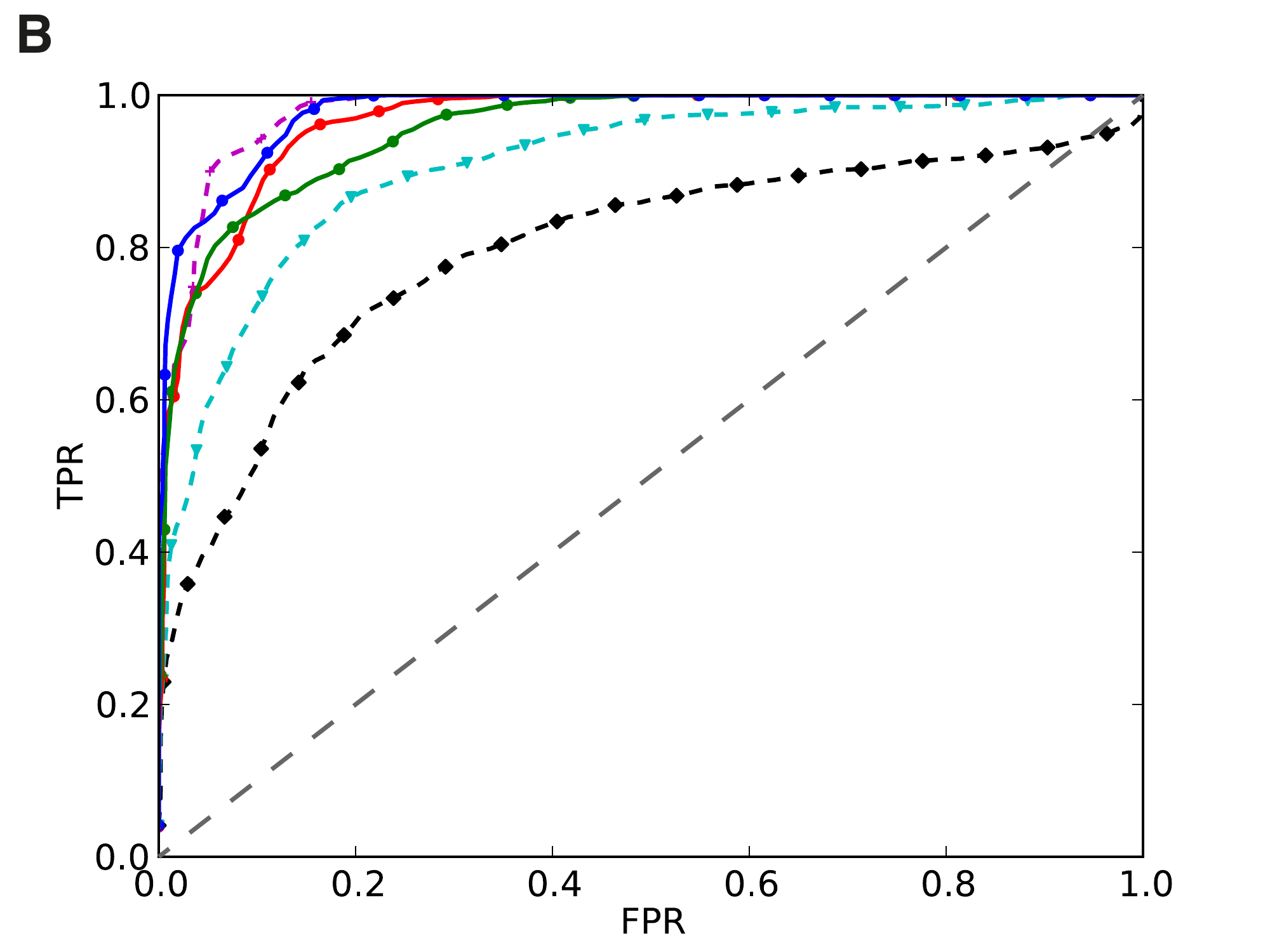}} \\
\resizebox{0.485\textwidth}{!}{\includegraphics*{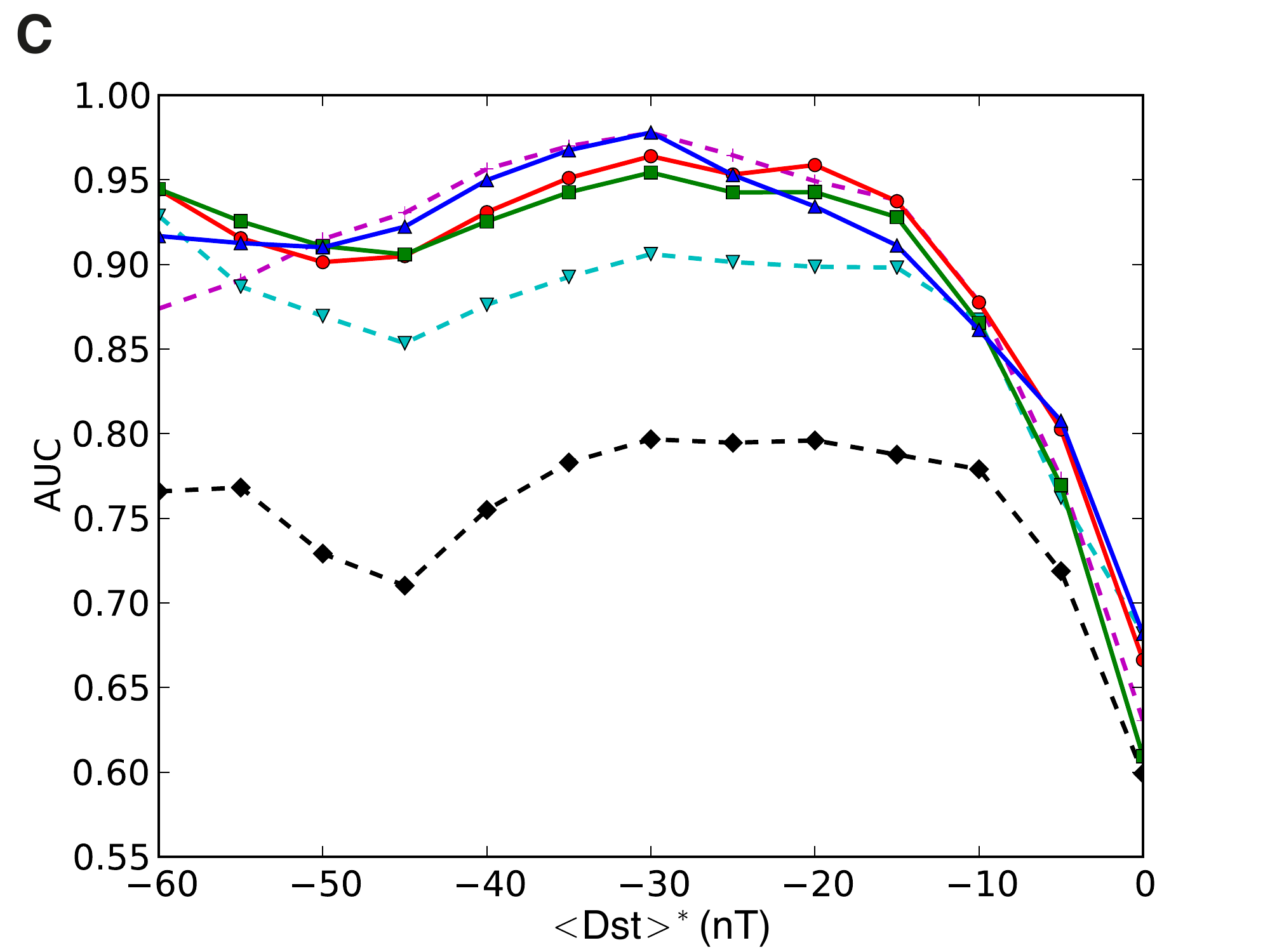}} \hfill
\resizebox{0.485\textwidth}{!}{\includegraphics*{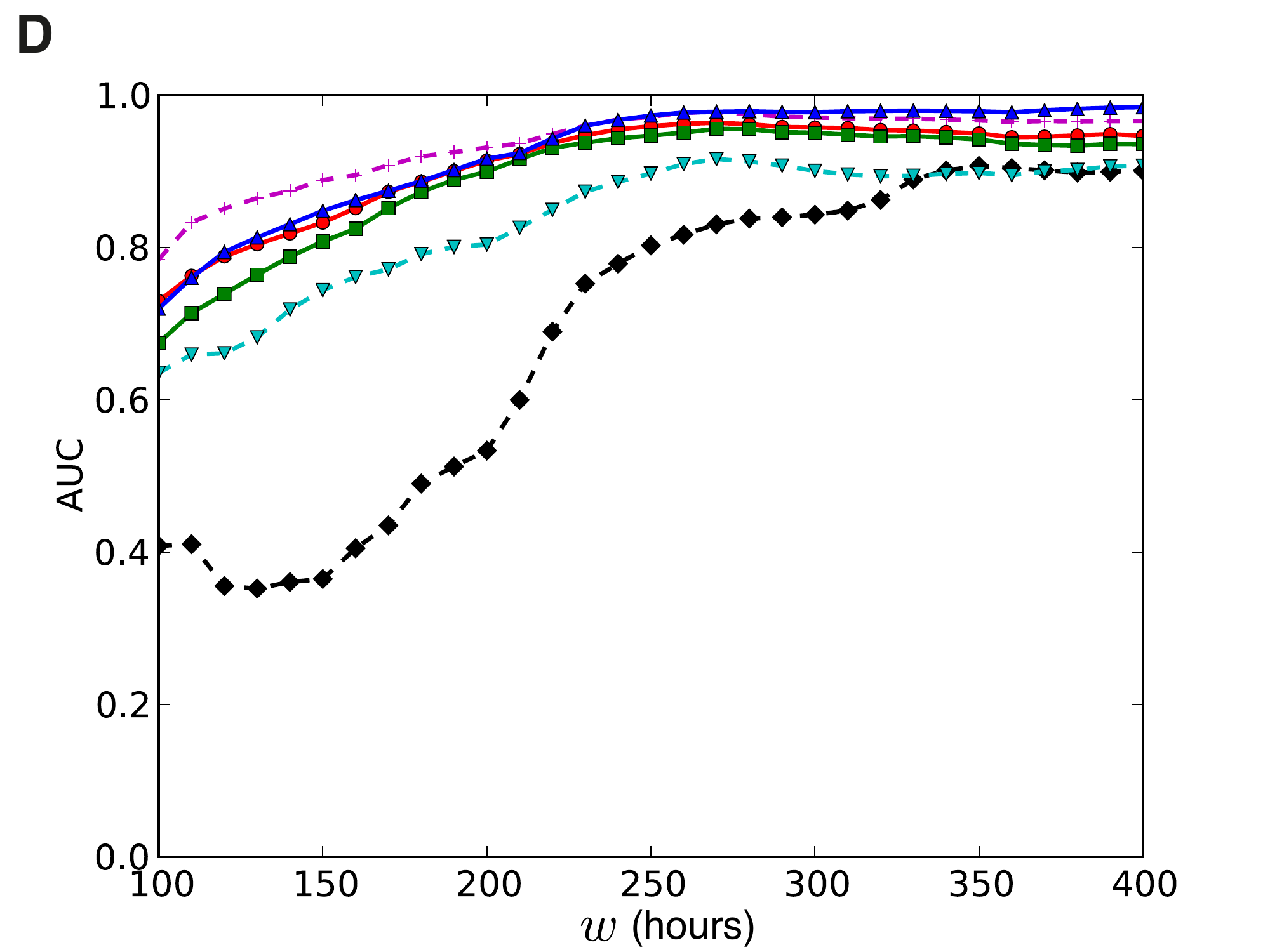}} \\
\caption{Results of ROC analysis for the RQA and RNA measures (line colors and symbols as in Fig.~\ref{fig:roc_fixed}) when using a data-adaptive discrimination between storm and non-storm conditions based on a threshold value $\left<\mbox{Dst}\right>^*$. (A) ROC curves using windows of $w=168$ hours, $\left<\mbox{Dst}\right>^*=-30$~nT. (B) As in (A) for $w=256$ hours. (C) Dependence of AUC on $\left<\mbox{Dst}\right>^*$ ($w=256$ hours), (D) AUC as a function of the window width $w$ ($\left<\mbox{Dst}\right>^*=-30$~nT).}
\label{fig:roc_recurrence}
\end{figure*}

Figure~\ref{fig:roc_recurrence}A,B shows the ROC curves of all six recurrence measures for two different choices of the window width $w$. In addition, we present the dependence of AUC on the discrimination threshold $\left<\mbox{Dst}\right>^*$ and running window width $w$, respectively, in Fig.~\ref{fig:roc_recurrence}C,D. 

When using the same window length of $w=256$ hours as in Sections~\ref{sec:performance} and \ref{sec:discrimination}, the comparative performance of the different characteristics follows the expectations based on the results of the correlation analysis in Section~\ref{sec:performance}. That is, $\mathcal{T}$, $TT$ and $DET$ generally provide the best discrimination between storm and non-storm conditions, followed by $LAM$, $\mathcal{C}$ and $\mathcal{L}$ (Fig.~\ref{fig:roc_recurrence}B). More specifically, with the variable threshold, $TT$ competes with $\mathcal{T}$ and $DET$ again, i.e., the performance characteristics are closer to the findings from correlation analysis than when using the heuristic coarse-grained classification into storm and non-storm periods from Section~\ref{sec:discrimination}. This difference is most likely caused by the dynamical properties of time windows with Dst values close to zero during the previously considered storm periods. Given the relatively high fraction of time windows with such conditions (see Fig.~\ref{fig:pdf_dst}), it is not unexpected that the six recurrence measures actually behave somewhat differently especially during such time intervals where the heuristic and data-adaptive classifications do not agree with each other. The potential relevance of this result -- particularly regarding the problem of anticipating or early detecting approaching magnetospheric disturbances from Dst index data -- could offer an interesting avenue for further research. However, for the latter purpose, additional information on solar wind parameters needs to be considered as well, bearing in mind that the magnetosphere is a driven system heavily affected by extraterrestrial forcing.

If we turn towards shorter windows (e.g., Fig.~\ref{fig:roc_recurrence}A), the general picture does not change much, but exhibits some interesting details: For shorter windows, the discrimination between storm and non-storm conditions becomes gradually worse for all measures, which is to be expected, since a lower amount of data is available for computing the measures the classification is based on. In turn, for longer windows, the classification skills of all recurrence parameters saturate at a window size between about 250 and 350 hours, depending on the specific measure (Fig.~\ref{fig:roc_recurrence}D). The latter scale of about 10 to 15 days appears related to the typical time-scales at which magnetospheric dynamics exhibits transitions between intense storm events and quiescence conditions.

In general, we find that for shorter windows the skills of $\mathcal{T}$ as expressed in terms of AUC decay somewhat slower than those of $TT$ and $DET$, making this measure (among the six characteristics studied in this work) most suitable for a temporally localized tracing of magnetospheric complexity variations. In turn, the behavior of $\mathcal{L}$ even changes qualitatively, providing AUC values below 0.5 (i.e., worse than a random classification, cf.~Fig.~\ref{fig:roc_recurrence}A,D). This observation is related to a transition from negative towards positive correlations between $\mathcal{L}$ and $\left<\mbox{Dst}\right>$ as $w$ decreases. This qualitative change in the behavior of $\mathcal{L}$ might be understood as follows: Larger windows typically cover a succession of different individual storm/quiescence intervals, indicating persistent changes in magnetospheric complexity with a two-state pattern within the same time window. In such a situation, we can expect high values of $\mathcal{L}$. In turn, shorter windows possibly only capture either one storm or non-storm phase, so that the recurrence characteristics relate to the dynamics of individual storms, which are more homogeneous and could therefore give rise to lower values of $\mathcal{L}$ than during (intermittent) short quiescence periods. In this spirit, both very high and very low values of $\mathcal{L}$ can be considered as indicators of regime changes (e.g., \cite{Donges2011PNAS}), a feature that has not yet been explored in full detail. Our present results provide indications towards a possible general explanation of this observational fact in terms of heterogeneity in the system's reconstructed phase space.

When varying the threshold $\left<\mbox{Dst}\right>^*$ for distinguishing between storm and non-storm conditions (Fig.~\ref{fig:roc_recurrence}C), we observe that all recurrence measures lose their skills when the classification becomes less informative (i.e., the threshold reaches values close to the normal background level of geomagnetic variations). In turn, for very high negative thresholds, there are only few storm periods remaining, so that the data available for classification become too sparse. As a reasonable trade-off, we recommend (for $w=256$ hours) an operational window of $\left<\mbox{Dst}\right>^*$ between about -40 and -20~nT, for which the AUC values are relatively stable, justifying our initial choice of a threshold at -30~nT.

\begin{figure}
\centering
\resizebox{0.485\textwidth}{!}{\includegraphics*{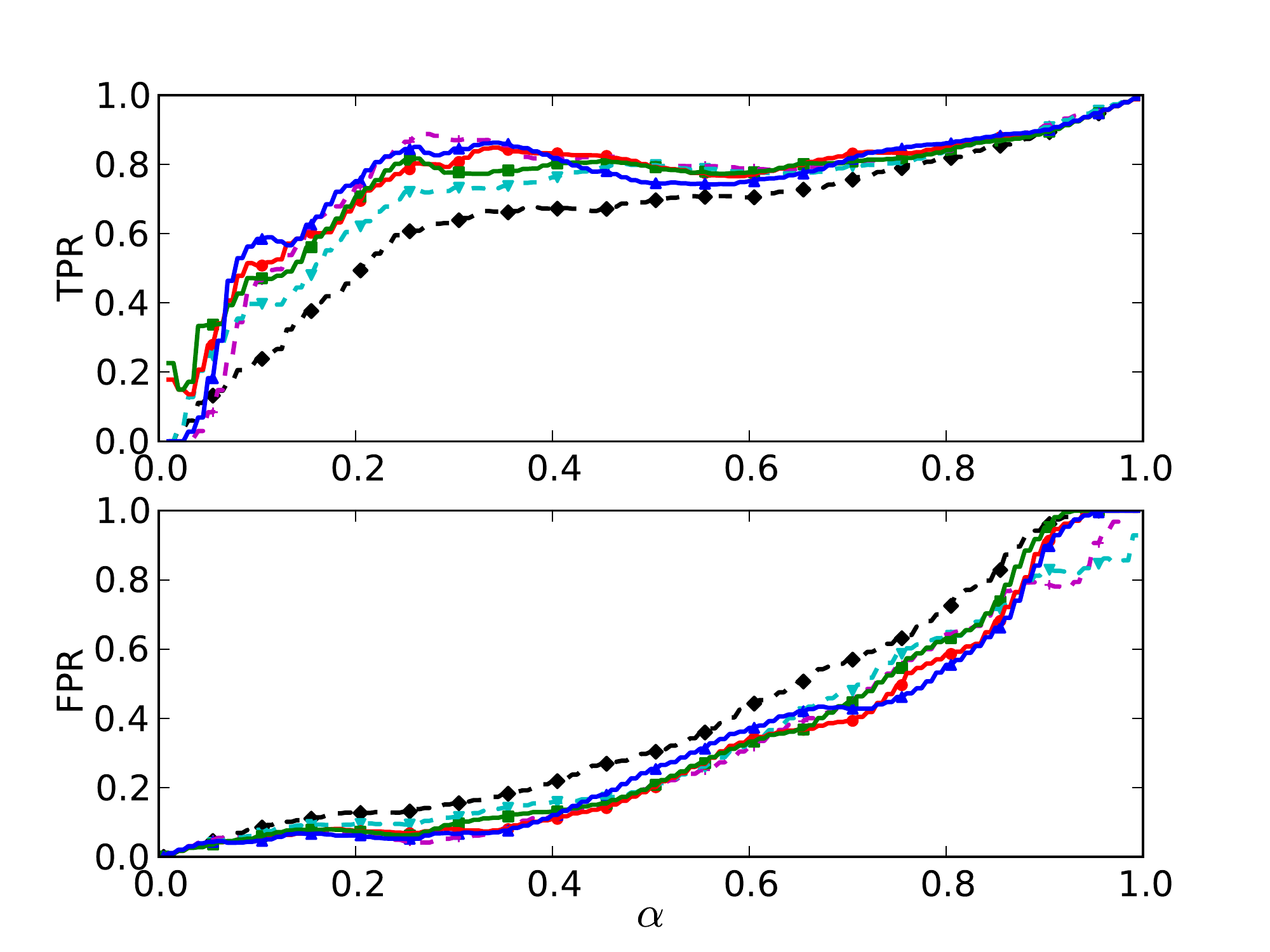}} 
\caption{True and false positive rates of detections for the six recurrence-based measures (symbols as in Fig.~\ref{fig:roc_recurrence}) when varying the threshold $\left<\mbox{Dst}\right>^*$ according to a certain quantile $\alpha$ of the distributions of $\left<\mbox{Dst}\right>$ and the associated quantile $1-\alpha$ of the respective recurrence measure (line colors and symbols as in Fig.~\ref{fig:roc_fixed}, $w=256$ hours, $\Delta w=1$ hour).}
\label{fig:tpr-fpr_recurrence}
\end{figure}

As a variant of the previous analysis, we finally modify the classical ROC approach by simultaneously considering thresholds for $\left<\mbox{Dst}\right>$ and the different recurrence characteristics. In order to ensure comparability, we fix the quantiles of the mean index value and the recurrence measure at the same values, which are then gradually varied. This implies that $\left<\mbox{Dst}\right>$ always exhibits equally many ``storm'' windows as the RQA/RNA measures. For each quantile $\alpha$, the true and false positive rates are monitored. The results of the corresponding analysis are shown in Fig.~\ref{fig:tpr-fpr_recurrence}. In general, the obtained patterns are compatible with the previously studied ROC curves: for small quantiles, both TPR and FPR are small, but rise towards higher quantiles. In general, the three RQA measures, $\mathcal{T}$ (for $\alpha\gtrsim 0.1$) and $\mathcal{C}$ (for $\alpha\gtrsim 0.4$) show the highest hit rates (TPR). In a similar way, the error rates (FPR) are commonly the lowest for $\mathcal{T}$, $TT$ and $DET$ for most quantiles. In turn, $\mathcal{L}$ displays clearly lower TPR and higher FPR than all other measures, again indicating that this property provides a less suited classifier for (window-mean) Dst values.

\subsection{Comparison with other methods}

\begin{figure*}
\centering
\resizebox{0.485\textwidth}{!}{\includegraphics*{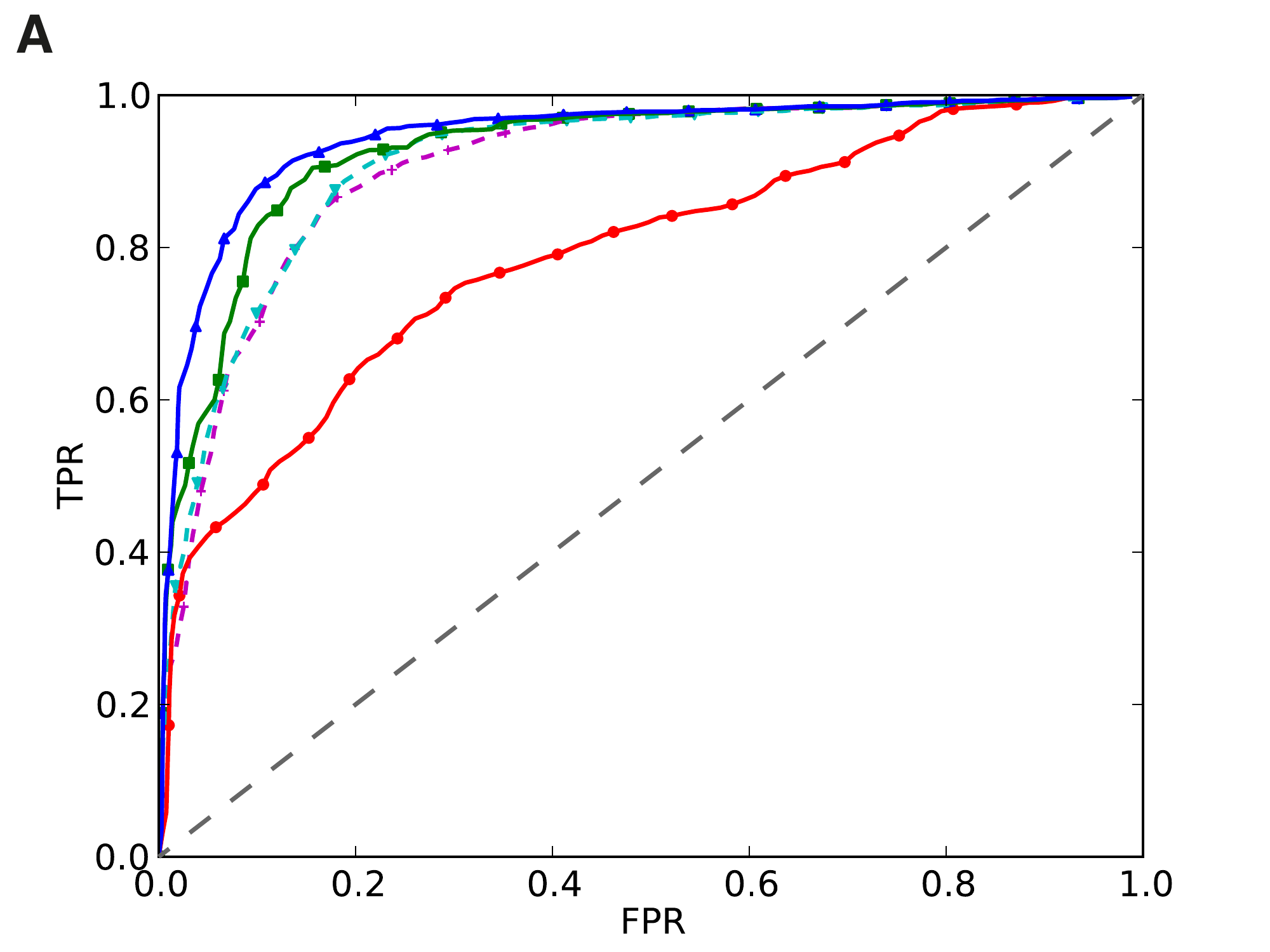}} \hfill
\resizebox{0.485\textwidth}{!}{\includegraphics*{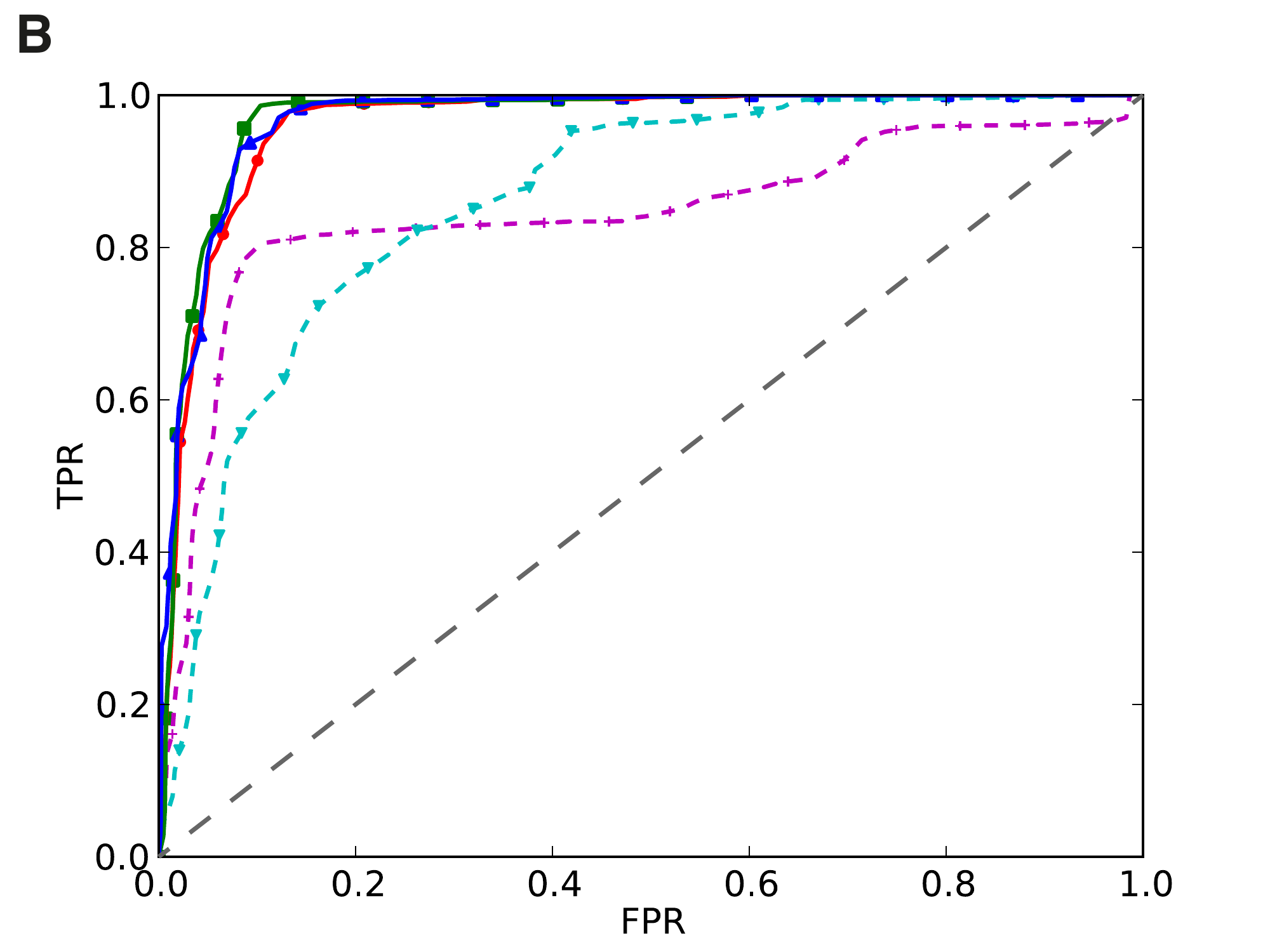}}  \\
\resizebox{0.485\textwidth}{!}{\includegraphics*{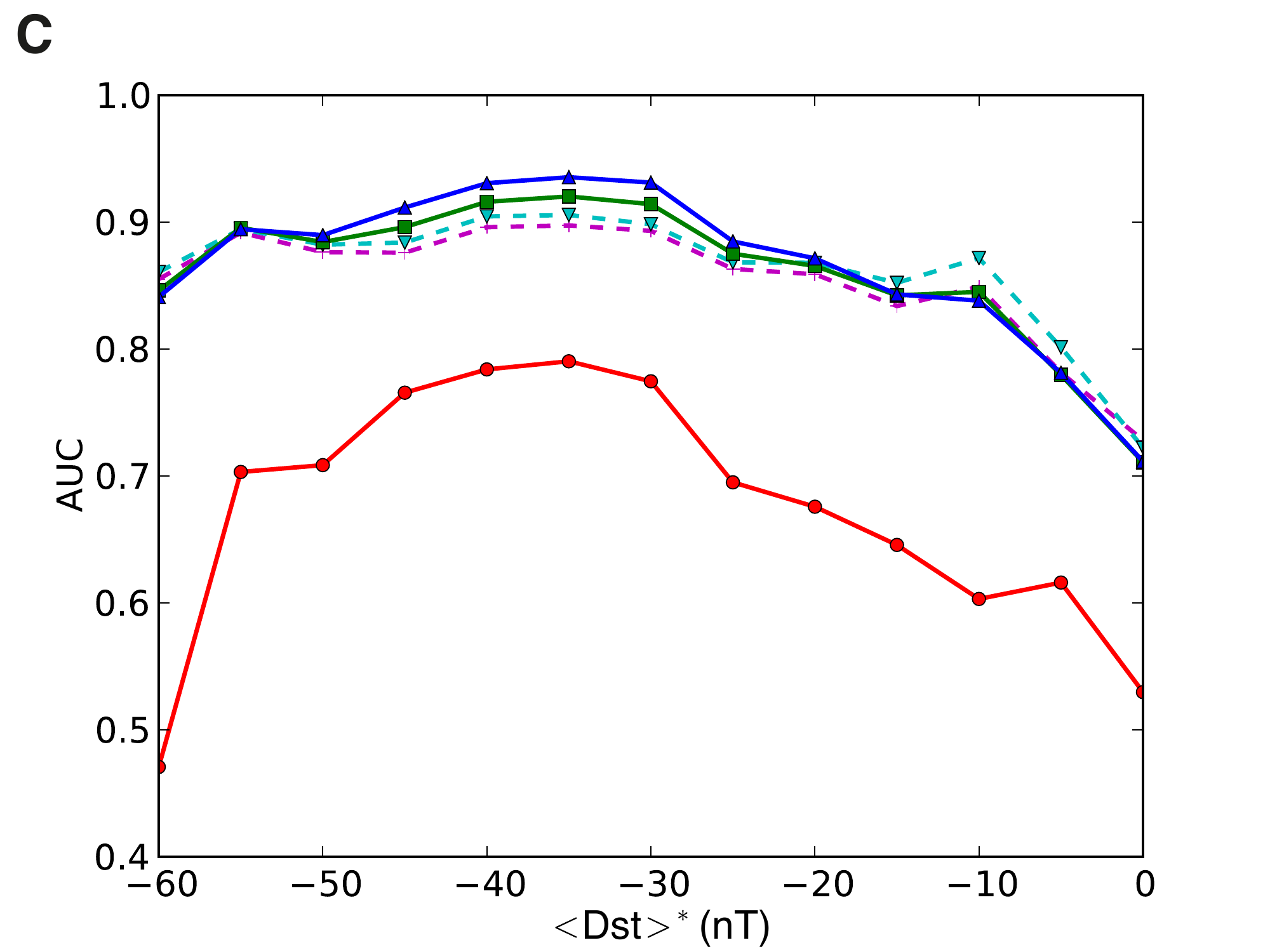}} \hfill
\resizebox{0.485\textwidth}{!}{\includegraphics*{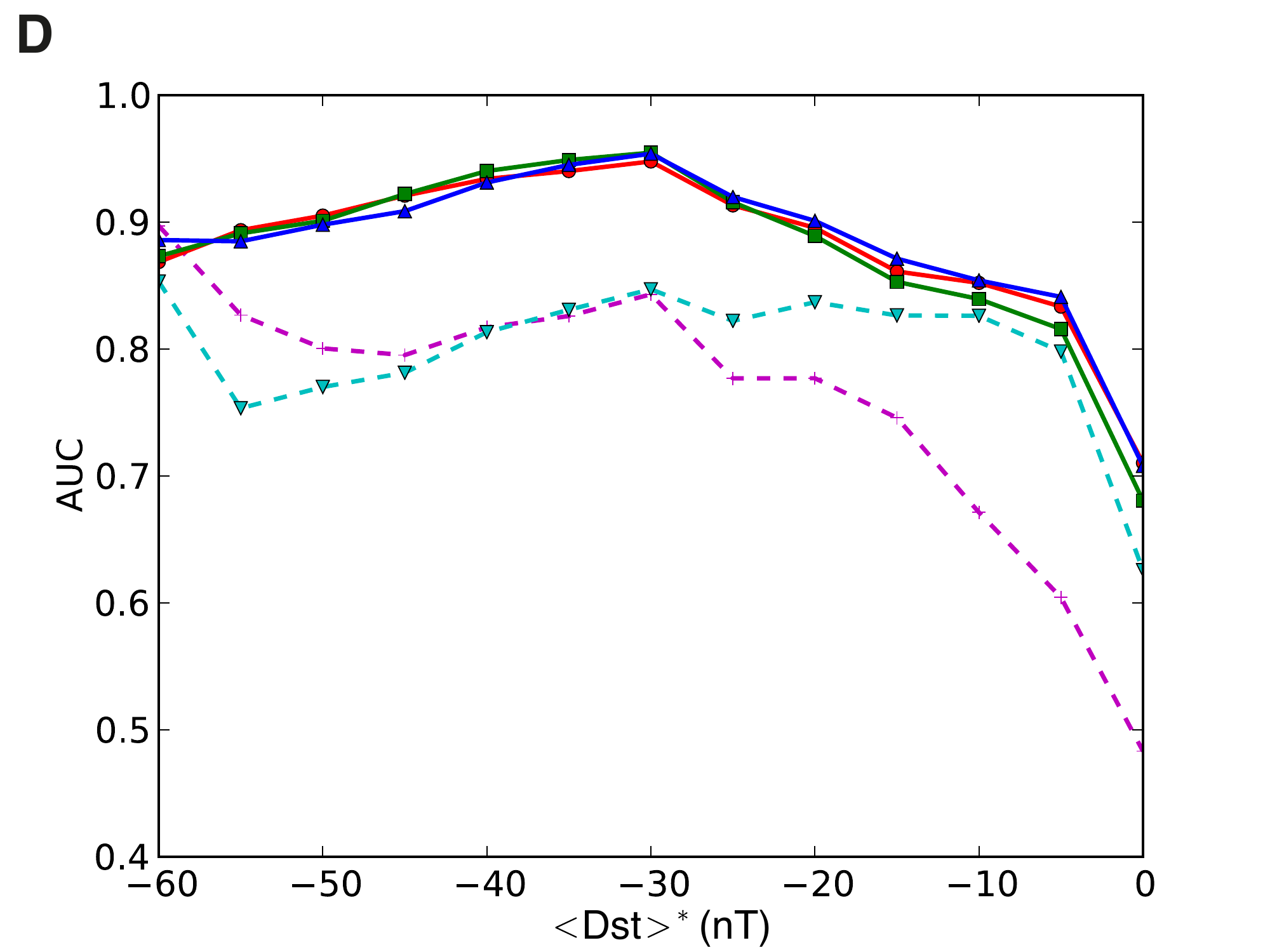}}
\caption{(A,B) As in Fig.~\ref{fig:roc_others_fixed}(A,B) for a data-adaptive classification with a fixed threshold $\left<\mbox{Dst}\right>^*=-30$~nT ($w=256$ hours, $\Delta w=1$ hour). (C,D) AUC values calculated from the ROC curves obtained for different values of $\left<\mbox{Dst}\right>^*$. Line colors and symbols as in (A,B).}
\label{fig:roc_others}
\end{figure*}

We repeat our ROC analysis for the non-recurrence characteristics previously discussed in Section~\ref{sec:discrimination}. Figure~\ref{fig:roc_others}A,B displays the resulting ROC curves, and Fig.~\ref{fig:roc_others}C,D the dependence of the resulting AUC values on the thresholds applied to the window-wise mean Dst values. Most of the studied characteristics again discriminate rather well between conditions with strongly negative versus close to zero Dst values. The observation that this feature is not unique to recurrence measures was to be expected given the previous results using entropic and correlation-based characteristics. 

In general, we again find the same three measures which perform worse than the others in the ROC analysis. In case of the Shannon entropy $S$, the relatively bad discriminatory skills are to be expected, since this measure does not characterize dynamical properties, but is calculated based on the (heavily coarse-grained) PDF of the Dst data within each window. For the Hurst exponent, we find that even for very high FRP, we do not reach TPR values sufficiently close to one, which indicates that there are certain storms where this measure takes values distinctively different from other storm periods. One possible reason for this could be the general problem of properly estimating $H$ from rather short time series segments \cite{Witt2013}. Finally, the LVD dimension density $\delta_{LVD}$ displays better discriminatory skills than $H$ and $S$, but distinguishes not as good as all considered dynamical entropies between storm and non-storm conditions. Note again that this measure is based on some linear approach, whereas entropic characteristics potentially account for nonlinearities. The imperfect discrimination between storm and non-storm periods by means of $\delta_{LVD}$ has already been reported by \cite{Donner2013} and could potentially be improved by systematically optimizing the parameters of this method. However, the latter measure has its particular advantage when considering very short windows, a setting where ``fully nonlinear'' characteristics commonly experience problems in their estimation.

Under ideal conditions, the three phase space based entropies AppEnt, SampEnt and FuzzyEnt provide the best skills in distinguishing storm from non-storm conditions, with AUC values of a similar order as for the best recurrence measures. This means that the observed differences between these entropies and some of the recurrence measures in case of the heuristic classification of storm and non-storm periods are largely relieved when considering a fully data-adaptive discrimination. In comparison with the phase space based entropies, the information-theoretic and statistical mechanics-based entropy characteristics exhibit slightly lower maximum AUC values, while the three aforementioned measures Shannon entropy, Hurst exponent and LVD dimension density generally display significantly lower values. For most measures, an optimum discrimination is again found for Dst threshold values of about -20 to -40~nT consistent with the corresponding results for the RQA/RNA measures.

\section{Discussion} \label{sec:discussion}

From the results described in the previous sections, we gain confidence that recurrence-based characteristics generally have great potentials in tracing temporal variations in the dynamical complexity of geomagnetic variations, but also other non-stationary geophysical systems. Regarding the inferred dynamical complexity of magnetospheric fluctuations during storm and non-storm conditions, the obtained results are in good agreement with the existing body of literature on this subject~\cite{Balasis2009}.

From the methodological perspective, the consideration of multiple measures from RQA and RNA in this study -- based on the same underlying structure, but characterizing different types of statistical properties -- allowed to discriminate between statistics that are better versus such that are less well suited for the purpose of distinguishing magnetic storms and quiescence phases based on the dynamical complexity of fluctuations of the Dst index. The obtained results provide interesting and (beyond the limits of this work) relevant information on the general potentials and applicability of these different characteristics. In turn, the fact that different recurrence measures appear to exhibit (at least quantitatively) different behaviors for transition periods between ``physiological'' and ``pathological'' states of the magnetosphere (in our case most notably short periods of relative quiescence of the magnetic field in between marked magnetic storms) makes such a multi-measure perspective a prospective approach for further studying magnetic field variations below general storm/quiescence variability, for example, regarding storm/substorm sequences. This aspect shall be further addressed in future work, thereby extending this study to both, additional time periods and other geomagnetic activity indices that are better suited for tracing other magnetospheric phenomena like substorms.

More specifically, the findings of this work demonstrate that some of the recurrence-based measures perform (depending on the specific setting) at least comparatively well as the best of the previously studied dynamical entropy characteristics in discriminating between storm and non-storm periods according to the dynamical complexity of magnetospheric variations. However, this tentative result should be further challenged from a methodological perspective: The so far best performing entropies have been developed based upon similar phase space based considerations as recurrence analysis and thus include the same intrinsic methodological parameters (i.e., embedding dimension $m$ and delay $\tau$). However, unlike for the recurrence measures used in this work, the optimal choice of these parameters has not yet been systematically addressed for these measures. Therefore, to this end we cannot provide a fair and finally conclusive comparison between entropies and recurrence characteristics, since a detailed study of different methodological settings for all entropy measures has been beyond the scope of this work. In general, we emphasize the potential of a further performance gain by systematically tuning embedding dimension and delay such as to achieve maximum AUC values.

From a conceptual viewpoint, the latter aspect raises the additional question of how to choose the embedding parameters in some optimum way in situations where the correlation properties of the data change crucially with time. For example, correlations in Dst values are much stronger expressed during magnetic storms than within periods of quiescence, as also indicated by the corresponding Hurst exponent and LVD dimension density \cite{Balasis2006,Donner2013}. In this work, we have chosen to fix the embedding delay at some globally justified value (selected according to commonly accepted standard criteria) disregarding the extremely strong temporal changes in the correlation structure. We emphasize that this approach ensures mutual comparability of the values of all dynamical characteristics obtained for different time windows and is thus advantageous for tracing variations in dynamical complexity, since recurrence measures (as well as other nonlinear dynamical properties) can exhibit a marked dependence on the embedding parameters. On the other hand, it could also be justified to take the opposite perspective and choose the embedding delay adaptively for each window. In fact, we have repeated the same recurrence analysis as shown in the previous sections using a time-dependent embedding delay, but did not obtain any conclusive results regarding the discrimination between dynamical complexity during storm and non-storm periods in that case (not shown).

Going even further into the methodological details of this study, another interesting aspect to be addressed in future work is to examine the different performance of the considered recurrence characteristics in more detail. For example, $TT$ and $LAM$ are based on the same rationale (i.e., quantifying statistical properties of the length distributions of vertical line structures in recurrence plots), but $TT$ clearly outperforms $LAM$. Following this observation, a promising approach could be considering the temporal changes associated with the full PDF of these line lengths to systematically address the question which statistical property (or combination of statistics) associated with this PDF provides the best discrimination between storm and non-storm conditions (and why). In the same way, one could proceed for the PDF of diagonal line lengths (e.g., comparing $DET$ with other characteristics like the mean diagonal line length $LMEAN$ not considered in this work) or different recurrence network properties based on transitive relationships (beyond $\mathcal{T}$ and $\mathcal{C}$). Yet another possible extension of the present work would be additionally considering recurrence time statistics (i.e., statistics based on the length of vertical non-recurrence structures or ``white lines'' in the recurrence plots, see, e.g., \cite{Ngamga2012}), which would provide a systematic extension to the study of return periods of storms and other magnetospheric disturbances towards  finer scales and more dynamical aspects of variability.

Furthermore, it should be noted that the AUC values reported in this work should be considered as estimates of the ``true'' values, which differ to a certain extent depending on the specific calculation strategy (in particular, the level of detail of the underlying ROC curves). In this spirit, given the finite length of the used data and the intrinsically non-stationary character of the Dst index, it might be useful to apply a further optimized estimation strategy. Such strategy could be based on computing the TPR and FPR values for each attained value of the respective measure in the full sample instead of some coarse-graining of the associated range or the PDF as used in this work. In addition, providing confidence bounds for the obtained AUC values (e.g., via cross-validation or bootstrapping techniques) would help statistically evaluating the differences between the skills of different methods in more detail. While such a treatment would have significantly enhanced the computational efforts of this study, given the multiple approaches used in this work, we are confident that the obtained results are at least qualitatively robust even without such further refinements.

Finally, we have demonstrated the robustness of our results for a given embedding dimension $m$ chosen according to the restrictions originating from our sliding window analysis and the associated window widths. In turn, it would be worth accounting for the possibly larger dimensionality of Dst index variations as suggested by the results in Fig.~\ref{fig:embedding}C,D, thereby contributing to the understanding of the dynamical behavior of the magnetosphere (respectively, the subsystem represented by Dst) as some low-dimensional dynamical system. Moreover, a systematic study of the sensitivity of our quantitative results with respect to the choice of the recurrence rate $RR$ would provide further information on characteristic ranges of Dst values during activity and quiescence periods of the magnetosphere (captured in terms of the associated recurrence threshold $\varepsilon$) which lead to qualitatively stable results of various flavours of recurrence analysis. Such further parameter studies have been, however, beyond the scope of the present work.

As emphasized above, all results obtained in this work are restricted to the properties of magnetospheric variability during one year of observations. From this analysis, we therefore cannot make any detailed statements about the generality of our findings for other years. Notably, not only the statistical distribution of storm and non-storm periods as well as storm magnitudes (as expressed by Dst variations or other geomagnetic indices) can be expected to vary from year to year (even beyond the solar Schwabe cycle). Even more, the specific characteristics of individual storm events may exhibit a certain range regarding the events' magnitudes, durations and dynamical characteristics, and differences in the latter aspect might result in modifications of our results. We expect such differences to be only quantitative rather than qualitative, but it would require a detailed investigation extending the data to the entire available period of observations to verify this expectation. Complying with other recent studies focusing on individual years of activity as well, we leave this aspect to be addressed in future work.

\section{Conclusions}  

We have employed a suite of selected characteristics from the modern toolbox of recurrence quantification analysis and recurrence network analysis to investigate the time-dependence of different aspects of dynamical complexity exhibited by the Dst index during one year with two marked periods of strong geomagnetic activity peaking in sequences of magnetic storms. While all considered measures have shown their ability to trace complexity variations associated with the succession of storm/non-storm periods very well, different characteristics exhibit different degrees of sensitivity with respect to changes in the magnetospheric variability patterns. Specifically, the recurrence network transitivity $\mathcal{T}$ -- together with two RQA measures -- has been identified as the most sensitive tracer of such variations. A detailed ROC analysis has shown that this property performs comparably well as (or even better than) the best dynamical entropy characteristics considered so far for the purpose of discriminating storm and non-storm conditions based on their nonlinear dynamical characteristics.

Our results provide new information that helps assessing and understanding the potentials of different measures emerging from this still quite novel approach. In this spirit, the fact that $\mathcal{T}$ performs particularly well could be due to this measure being directly related to a generalized notion of fractal dimension \cite{Donner2011EPJB} associated with the geometric structure of the data in the phase space reconstructed by means of time-delay embedding. Besides other recent applications of the same characteristic \cite{Donges2011,Donges2011PNAS,Donges2014}, this study is among the few cases where fractal dimension concepts have been successfully used for tracing temporal variations of the dynamical complexity of geophysical systems based on single time series. This observation opens promising new research avenues by revisiting the classical concept of fractal dimensions in a geoscientific context.

Regarding the temporal organization of fluctuations inside the Earth's magnetosphere, our findings confirm previous results on the distinctive difference between magnetic storms and quiescence periods. Specifically, storm periods exhibit an elevated degree of dynamical regularity related to the gradual trends of the Dst index during the emergence of magnetic storms and the subsequent recovery phase. The multiplicity of recurrence measures studied in this work allows capturing different facets of the dynamical complexity. Specifically, RQA measures are explicitly linked with the dynamical organization of the system's fluctuations in the sense that they highlight persistent proximity relationships between the embedded values of the Dst index in the associated reconstructed phase space. In turn, RNA measures capture geometric properties of the multi-dimensional distributions of these state vectors and, thus, rather take a structural perspective on that space. The fact that measures from both classes (particularly $\mathcal{T}$ versus $TT$ and $DET$) exhibit very similar temporal variations underlines that both aspects are closely entangled in the magnetosphere. Further addressing this fact by considering other variability indicators based on different rationales than Dst could therefore allow more detailed insights into the dynamical complexity of the magnetosphere associated with different spatial and temporal scales and processes. In turn, this could potentially advance operational models used for space weather forecasting purposes.

\section*{Acknowledgments}
This work has been financially supported by the joint Greek-German project ``Transdisciplinary assessment of dynamical complexity in magnetosphere and climate: A unified description of the nonlinear dynamics across extreme events'' funded by IKY and DAAD. Individual financial support of the authors has been granted by the LINC (Learning about Interacting Networks in Climate) project (project no. 289447) funded by the Marie Curie Initial Training Network (ITN) program (FP7-PEOPLE-2011-ITN), the German Federal Ministry for Science and Education (BMBF) via the Young Investigator's Group CoSy-CC$^2$ (grant no. 01LN1306A) and the project GLUES, the Stordalen Foundation (Planetary Boundary Research Network PB.net), and the International Research Training Group IRTG 1740/TRP 2014/50151-0, jointly funded by the German Research Foundation (DFG, Deutsche Forschungsgemeinschaft) and the S\~{a}o Paulo Research Foundation (FAPESP, Funda\c{c}\~{a}o de Amparo \`a Pesquisa do Estado de S\~{a}o Paulo). Numerical codes used for estimating RQA and RNA properties can be found in the software package pyunicorn \cite{Donges2013}, which is available at \url{https://github.com/pik-copan/pyunicorn}. The Dst data have been obtained from the World Data Center for Geomagnetism, Kyoto (\url{http://wdc.kugi.kyoto-u.ac.jp/index.html}). We are grateful to three reviewers of an earlier version of this manuscript for their detailed comments.

\bibliographystyle{unsrt}
\bibliography{library_v2}

\end{document}